\documentclass[preprint,prd,aps,superscriptaddress,11pt]{revtex4-1}

\usepackage{graphicx}
\usepackage{amsmath}
\usepackage{amssymb}
\usepackage{dcolumn}
\usepackage{bm}
\usepackage{slashed}
\usepackage{color}
\usepackage{changepage}
\usepackage{multirow}
\usepackage{hyperref}
\usepackage{accents}

\newcommand{\ubar}[1]{\underaccent{\bar}{#1}}
\newcommand{\be}{\begin{equation}}
\newcommand{\ee}{\end{equation}}
\newcommand{\ba}{\begin{eqnarray}}
\newcommand{\ea}{\end{eqnarray}}
\newcommand{\no}{\nonumber \\}
\newcommand{\gsim}{\mathrel{\hbox{\rlap{\lower.55ex \hbox {$\sim$}}
                   \kern-.3em \raise.4ex \hbox{$>$}}}}
\newcommand{\lsim}{\mathrel{\hbox{\rlap{\lower.55ex \hbox {$\sim$}}
                   \kern-.3em \raise.4ex \hbox{$<$}}}}

\def\hd{{\hat \d}}
\def\roughly#1{\mathrel{\raise.3ex\hbox{$#1$\kern-.75em%
\lower1ex\hbox{$\sim$}}}}
\def\lsim{\roughly<}
\def\gsim{\roughly>}

\def\vy{{\vec y}}
\def\({\left(}
\def\){\right)}
\def\[{\left[}
\def\]{\right]}
\def\<{\langle}
\def\>{\rangle}

\def\bik{{\textbf {\textit k}}}
\def\bix{{\textbf {\textit x}}}
\def\tcO{\tilde{\cal O}}

\newcommand{\myquad}[1][1]{\hspace*{#1em}\ignorespaces}

\def\k{{\kappa}}
\def\l{{\lambda}}
\def\L{{\Lambda}}
\def\d{{\delta}}
\def\D{{\Delta}}
\def\o{{\omega}}

\def\e{{\epsilon}}

\def\a{{\alpha}}
\def\b{{\beta}}
\def\c{{\chi}}

\def\g{{\gamma}}
\def\G{{\Gamma}}
\def\p{{\pi}}

\def\m{{\mu}}
\def\n{{\nu}}
\def\r{{\rho}}
\def\s{{\sigma}}
\def\S{{\Sigma}}
\def\t{{\tau}}
\def\th{{\theta}}
\def\Th{{\Theta}}
\def\ph{{\phi}}

\def\x{{\xi}}
\def\et{{\eta}}
\def\z{{\zeta}}

\def\hb{{\hbar}}

\def\ep{{\varepsilon}}
\def\vrp{{\varpi}}

\def\cA{{\cal A}}
\def\cB{{\cal B}}
\def\cO{{\cal O}}
\def\cR{{\cal R}}
\def\cS{{\cal S}}
\def\cT{{\cal T}}

\def\cV{{\cal V}}

\def\cN{{\cal N}}

\def\ms{{\mathring s}}
\def\mS{{\mathring \S}}
\def\mTh{{\mathring \Th}}

\def\cTh{{\check \Th}}
\def\chS{{\check \S}}
\def\chs{{\check s}}

\def\ue{{\ubar \ep}}
\def\eb{{\bar \ep}}
\def\uR{{\ubar \cR}}
\def\uB{{\ubar \cB}}
\def\bT{{\bar T}}
\def\uT{{\ubar T}}
\def\pb{{\bar p}}
\def\up{{\ubar p}}
\def\uv{{\ubar v}}
\def\upi{{\ubar \p}}

\newcommand{\fr}{\frac}

\newcommand{\pd}{\partial}

\newcommand{\pr}{\parallel}

\date{\today}

\begin{document}

\title{\bf Relativistic spin hydrodynamics revisited with general rotation by entropy-current analysis}%of Dirac fermions
\author{Lixin Yang}
\email{yanglixin@fudan.edu.cn}
\affiliation{Institute of Modern Physics, Fudan University, Shanghai 200433, China}
\author{Li Yan}
\email{cliyan@fudan.edu.cn}
\affiliation{Institute of Modern Physics, Fudan University, Shanghai 200433, China}

\begin{abstract}

We revisit the canonical formulation of spin hydrodynamics for Dirac fermions with a general thermal vorticity. The orders of the general thermal vorticity and the corresponding spin variables are considered independently from those of the conventional hydrodynamic variables and their perturbative gradients. Assuming a totally antisymmetric spin current of Dirac fermions, the entropy-current analysis with a general spin potential indicates that the constitutive relations of the stress-energy tensor have to involve spin variables, particularly those linked to boost symmetry, to adhere to the entropy principle. In the presence of the degree of freedom associated with boost symmetry, we choose the constitutive relations of the canonical formulation to be connected to those of the phenomenological formulation through pseudogauge transformation. Subsequently, a linear-mode analysis is conducted using the resulting spin hydrodynamic equations. It is observed that the spin and hydrodynamic modes in this canonical formulation display different characteristics compared to those in the phenomenological formulation up to the second order of gradient.

\end{abstract}

\maketitle

%\vspace{0.1in}

\newpage

\section{Introduction}

Relativistic hydrodynamics, as an effective theory in terms of the IR variables, has proven highly successful in describing the macroscopic behavior of various many-body systems, spanning from astrophysics to relativistic heavy-ion collisions. Spin-orbit coupling plays a significant role in relativistic fluids with spinful constituents, leading to spin polarization, particularly in the presence of substantial angular momentum and/or strong vorticity. The polarization phenomena have been intensively studied in heavy-ion collisions since long \cite{Liang:2004ph,Liang:2004xn,Voloshin:2004ha,Betz:2007kg,Becattini:2007sr,Huang:2011ru}, and its occurrence has been confirmed through experimental measurements of hyperon polarization \cite{STAR:2017ckg,STAR:2018gyt,STAR:2020xbm} and vector-meson spin alignment \cite{ALICE:2019aid,STAR:2022fan,ALICE:2022dyy}. The global polarization of $\L$ hyperons has been effectively captured by relativistic hydrodynamic models incorporating thermalized spin degrees of freedom \cite{Becattini:2007sr,Becattini:2007nd,Becattini:2013vja,Becattini:2013fla,Becattini:2015ska,Becattini:2016gvu,Karpenko:2016jyx,Pang:2016igs,Xie:2017upb}. However, theoretical calculations \cite{Becattini:2017gcx,Becattini:2020ngo} concerning the azimuthal-angle dependence of hyperon polarization have yielded results with opposite signs compared to the experimental data \cite{Niida:2018hfw,STAR:2019erd}. This discrepancy, known as the``sign problem" in explaining local spin polarization, has also been addressed in related reviews \cite{Liu:2020ymh,Gao:2020vbh,Huang:2020xyr,Becattini:2022zvf}.
%This problem is currently the focus of intense work \cite{Florkowski:2019qdp,Florkowski:2019voj,Zhang:2019xya,Becattini:2019ntv,Xia:2019fjf,Wu:2019eyi,Sun:2018bjl,Liu:2019krs}, however, a convincing solution does not yet exist.
%In order to address this problem one needs to understand how the orbital angular momentum of the strongly interacting matter created in noncentral heavy-ion collisions is converted into the spin angular momentum of its constituents. In order to account for the nontrivial dynamics of the spin degrees of freedom, it has been proposed to introduce the rank-three spin tensor as additional dynamical variable, promoting relativistic hydrodynamics to a theory of spin hydrodynamics \cite{Florkowski:2017ruc,Florkowski:2017dyn,Florkowski:2018myy,Becattini:2018duy,Florkowski:2018fap,Hattori:2019lfp,Bhadury:2020puc}
%constitutive relations \cite{Denicol:2012cn,Molnar:2013lta,Denicol:2014vaa}

Relativistic spin hydrodynamics is a promising framework for understanding the sign problem, which has been advancing rapidly through various approaches in a significant body of research \cite{Montenegro:2017rbu,Montenegro:2017lvf,Florkowski:2017ruc,Florkowski:2018fap,Montenegro:2018bcf,Hattori:2019lfp,Li:2019qkf,Montenegro:2020paq,Garbiso:2020puw,Gallegos:2020otk,Fukushima:2020ucl,Bhadury:2020puc,Li:2020eon,Shi:2020htn,Hu:2021lnx,Gallegos:2021bzp,Peng:2021ago,Hongo:2021ona,Wang:2021ngp,Wang:2021wqq,She:2021lhe,Hu:2021pwh,Cartwright:2021qpp,Hongo:2022izs,Weickgenannt:2022zxs,Bhadury:2022ulr,Cao:2022aku,Daher:2022xon,Singh:2022ltu,Gallegos:2022jow,Biswas:2023qsw,Xie:2023gbo,Becattini:2023ouz}. Essentially, relativistic spin hydrodynamics is constructed by introducing the spin modes linked to Lorentz symmetry as additional IR variables, alongside the conventional hydrodynamic variables related to relativistic translation invariance. The spin variables can typically be categorized into rotation and boost components based on the subgroups of the Lorentz transformation. The former characterizes the spin polarization in the fluid, while the role of the latter remains unclear. An entropy-current analysis approach to spin hydrodynamics was introduced in \cite{Hattori:2019lfp} with the canonical spin current being antisymmetric only in its last two indices, referred to as the phenomenological formulation. The same approach is also implemented in the canonical formulations \cite{Hongo:2021ona,Cao:2022aku} where the canonical spin current of Dirac fermions is totally anti-symmetric. It is noteworthy that the canonical formulations in \cite{Hongo:2021ona,Cao:2022aku} are developed without considering the degree of freedom associated with boost symmetry. In contrast, another canonical formulation incorporating boost variables is presented in \cite{Daher:2022xon}, which is connected to the phenomenological formulation through a pseudogauge transformation. It is meaningful to explore the inclusion of boost variables in a general relativistic framework, since boost symmetry, along with rotation, is a fundamental aspect of Lorentz covariant hydrodynamics. Furthermore, hydrodynamics with varying approaches to the treatment of boost modes may exhibit distinct behavior when subjected to spin-orbit coupling. 

In this paper, we investigate the canonical formulation of spin hydrodynamics by considering its applicability to spinful fluids across a broad range of thermal vorticity intensities. In this scenario, the spin potential is presumed to be a general antisymmetric tensor in order to coincide with the thermal vorticity in equilibrium. Moreover, the magnitudes of both the thermal vorticity and the spin potential are assessed without regard to perturbative gradients. In the case of general spinful fluids, the entropy-current analysis suggests that the entropy principle cannot be fulfilled unless spin variables are included in the constitutive relations of the stress-energy tensor, with these spin variables needing to account for the degree of freedom linked to boost symmetry. The constitutive relations in the presence of boost variables have not yet been definitively determined. For simplilcity, we opt to connect the canonical formulation of spin hydrodynamics to the phenomenological approach through pseudogauge transformation. The linear-mode analysis using the resulting spin hydrodynamic equations reveals that the spin and hydrodynamic modes in this canonical formulation exhibit distinct dispersion relations compared to the phenomenological formulation up to the second order of gradients.

%The paper is organized as follows: In Sec.\eqref{SecDrc}, we review the canonical formulation of spin hydrodynamics for Dirac fermions in the entropy-current analysis approach. By utilizing the resulting constraints on the constitutive relations, we demonstrate that the entropy principle can not be satisfied in the absence of the boost variables in Sec.\eqref{SecIss}. By including the degree of freedom linked to boost symmetry, we give the general constraint on the constitutive relations of the canonical formulation in Sec.\eqref{SecConstnt}. We then choose the solution to this constraint to correspond to the a pseudogauge transformation from the phenomenological formulation and perform the linear-mode analysis in Sec.\eqref{SecLnr}. 

Throughout this paper, we adopt the mostly plus Minkowski metric $\et^{\m\n}\equiv$diag$\(-1, 1, 1, 1\)$. For the Levi-Civita symbol, we use the convention $\e^{0123}=-\e_{0123}=1$ and $\e^{123}=\e_{123}=1$. We also define the notations $X^{(\m\n)}\equiv \fr{1}{2}\(X^{\m\n}+X^{\n\m}\)$ and $X^{[\m\n]}\equiv \fr{1}{2}\(X^{\m\n}-X^{\n\m}\)$.

\section{Spin hydrodynamics for Dirac fermions}\label{SecDrc}

The conservation equations of the Noether's currents from the relativistic translation and Lorentz symmetry are
\begin{align}
&\pd_\m\Th^{\m\n}=0,\label{Em}\\
%&\pd_\m\S^{\m\n\a}=\Th^{\a\n}-\Th^{\n\a}.
&\pd_\a J^{\a\m\n}=\pd_\a\S^{\a\m\n}+\Th^{\m\n}-\Th^{\n\m}=0,\label{Am}%\label{pt}
\end{align}
where the total angular momentum is $J^{\a\m\n}\equiv \S^{\a\m\n} + x^\m\Th^{\a\n}-x^\n\Th^{\a\m}$ with $\S^{\a\m\n}$ being the spin tensor. The canonical stress-energy tensor $\Th^{\m\n}$ is asymmetric in its two indices, comprising both symmetric and antisymmetric components, while the total angular momentum density $J^{\a\m\n}$ is antisymmetric in its last two indices. 

In classical physics, the hydrodynamics of the Quark-Gluon Plasma (QGP) is formulated with IR variables while the details at a microscopic level are averaged out. For simplicity, we consider spin hydrodynamics without conserved charges. % \cite{Liu:2018kfw}  $\cR^{\m\n}$, and consequently, in  we embrace the perspective of effective field theory, which posits that the UV variables average out, with their effects encoded in the IR variables. Our focus is on the quarks fluid, where effects of gluons are Although the coarse-graining process may prompt further questions about how the decomposition of the total angular momentum is impacted, 
We follow \cite{Hongo:2021ona} to assume that the coarse-grained spin tensor in hydrodynamics retains the entire antisymmetry of its corresponding quantum operator. %, which amounts to postulating that the spin effects from gluons are integrated out as microscopic details while leaving the total antisymmetry of the spin tensor intact. 
Employing the fluid four-velocity $u^\m$ with $u^\m u_\m=-1$, the spin density is introduced as $\cR^{\m\n}\equiv-u_\a\S^{\a\m\n}$ with $\cR^{\m\n}=-\cR^{\n\m}$. 
The resulting spin density satisfies $\cR^{\m\n}u_\n=0$ due to the total antisymmetry of $\S^{\m\n\a}$. As a result, $\cR^{\m\n}$ is fully saturated with only three independent components associated with the spatial rotation symmetry, while the remaining three attached to the boost symmetry are absent in the spin tensor $\S^{\m\n\a}$\footnote{In general, a totally antisymmetric rank-3 tensor $\S^{\a\m\n}$ can contain, at most, four independent components, of which only three can be present in $\cR^{\m\n}$ by the definition $\cR^{\m\n}\equiv-u_\a\S^{\a\m\n}$. In any case, the totally antisymmetric spin tensor cannot encompass all six independent components associated with Lorentz symmetry.}. 
In such condition, it is yet to be determined whether the spin variables, especially the boost ones, can be generally absent in the coarse-grained stress-energy tensor. To this end, we perform an entropy-current analysis to constrain the presence of spin variables in hydrodynamics with the entropy principle. 
%It is beyond the scope of this work to figure out how the coarse-graining casts a quantum framework into its IR counterpart. It is worth noting that in the QCD angular momentum tensor from quantum gauge theory, there may exist a boost part and a gluon part alongside the quark part, as discussed in \cite{Wakamatsu:2010cb}. To ensure the total antisymmetry of the spin tensor in spin hydrodynamics, After all, boost and rotation are of equal status in Lorentz transformation as essential ingredients In a Lorentz covariant framework that has promoted    Given that the degree of freedom of spin fields under rotation can survive the coarse-graining process to emerge as spin modes, it is conceivable that the boost part could be incorporated into the stress-energy tensor and produce slow variables after the integration of fast variables. Constructing spin hydrodynamics from a first-principle approach, commencing from quantum gauge theory, . 
%and that spin polarization represents an additional independent dynamical mode within the realm of spin hydrodynamics. Employing the fluid four-velocity $u^\m$, we introduce the antisymmetric spin density $\cR^{\m\n}\equiv-u_\a\S^{\a\m\n}$. 
%of spin density attached to the spatial rotation symmetry there are only three independent components in the spin density 

%We will perform an entropy-current analysis We denote by $\cB^{\m\n}$ with $\cB^{\m\n}=-\cB^{\n\m}$ and $\cB^{\m\n}u_\n\neq 0$. 
The totally antisymmetric spin tensor can be decomposed into longitudinal and transverse parts as
\begin{align}
\S^{\m\n\a}=u^{\m} \cS^{\n\a} + u^{\n} \cS^{\a\m} + u^{\a} \cS^{\m\n} + \e^{\m\n\a\s}u_\s\mS,\nonumber
\end{align}
where the antisymmetric component $\cS^{\m\n}$ can be further decomposed as
\begin{align}
\cS^{\m\n}=\e^{\m\n\r\s} \cR_\r u_\s + 2u^{[\m}\cB^{\n]},\nonumber
\end{align}
with $\cR_\a=\fr{1}{2}\e_{\a\m\n\s}\cS^{\m\n}u^\s$ and $\cB^\m=\cS^{\m\n}u_\n$. 
%with $\cR\cdot u=\cB\cdot u=0$ and $\cR,\,\cB$ being an axial vector and a polar vector respectively. 
Noting the identities 
\begin{align}
\e^{\m\n\a\s}\cR_\s &= \(u^{\m} \e^{\n\a\r\s} + u^\n\e^{\a\m\r\s} + u^\a\e^{\m\n\r\s}\) \cR_\r u_\s ,\no
0 &= u^{\m} u^{[\n} \cB^{\a]} + u^\n u^{[\a} \cB^{\m]} + u^\a u^{[\m} \cB^{\n]} , \label{IDcS}
\end{align}
one readily writes the spin tensor into
\begin{align}
\S^{\m\n\a}=\e^{\m\n\a\s}\(\cR_\s + u_\s\mS\),%\nonumber
\end{align}
which immediately gives $\cR^{\m\n}=\e^{\m\n\r\s}\cR_\r u_\s$. Since $\cR_\r$ captures all the three independent components of $\cR^{\m\n}$ in a covariant form, $\mS$ is automatically left as corrections out of equilibrium to be constrained by the entropy principle. Given $\cR^{\m\n}$ representing the rotation components of the spin modes $\cS^{\m\n}$, $\cB^{\m\n}=2u^{[\m}\cB^{\n]}$ as the rest part naturally denotes the boost modes where $\cB^\m$ contains all the three independent components related to the boost symmetry. %$S=R+B$ in the last section which decomposes the spin part of the Lorentz generators into rotation and boost components
\iffalse
To be concrete, based on the decomposition in \eqref{decpAm}, it is natural to explicitly decompose $\cS^{\m\n}$ and $\o_{\m\n}$ as 
Utilizing the decomposition in \eqref{decpAm}, it is appropriate to explicitly decompose $\cS^{\m\n}$ and $\o_{\m\n}$ as
\begin{align}
&\cS^{\m\n}=\cR^{\m\n} + \cB^{\m\n}=\e^{\m\n\a\s} \cR_\a u_\s + 2u^{[\m}\cB^{\n]},\no
&\o_{\m\n}= r_{\m\n} + b_{\m\n}=\e_{\m\n\a\s} r^\a u^\s + 2u_{[\m} b_{\n]} ,
\end{align}
with $\cB^\m\equiv \cB^{\m\n}u_\n$, $b_\m\equiv b_{\m\n}u^\n$ and $r^\s\equiv \fr{1}{2}\e^{\s\a\m\n}u_\a r_{\m\n}$. We note that $\cR^\m u_\m=\cB^\m u_\m=r_\m u^\m=b_\m u^\m=0$. Therefore, $\cR^\m$ exactly captures all the three independent components of $\cR^{\m\n}$. The cases are analogous for the relations of $\cB^\m$, $r_\m$ and $b_\m$ to $\cB^{\m\n}$, $r_{\m\n}$ and $b_{\m\n}$, respectively. %  we have cast into the six degrees of freedom of $\cS^{\m\n}$ into three of $\cR_\a$ attached to the spatial rotation symmetry and the other three of $\cB^\m$ to the boost symmetry. 
Noting the identities $\e^{\m\n\a\s}\cR_\s = u^{\m} \cR^{\n\a} + u^\n\cR^{\a\m} + u^\a\cR^{\m\n} $ and $0 = u^{\m} \cB^{\n\a} + u^\n\cB^{\a\m} + u^\a\cB^{\m\n} $,
\fi

The local thermodynamic relations generalized with spin variables are 
\begin{align}
&T s=\ep+p-\fr{1}{2}\o_{\m\n}\cS^{\m\n},\label{Euler}\\%\qquad\text{Euler equation}
&T ds=d\ep-\fr{1}{2}\o_{\m\n}d\cS^{\m\n},\label{firstlaw}\\
&s dT=dp-\fr{1}{2}\cS^{\m\n}d\o_{\m\n},\label{GibDuh}%\qquad\text{Gibbs-Duhem equation}
\end{align}
where $T$, $s$, $\ep$, $p$ and $\o_{\m\n}=-\o_{\n\m}$ denoting the local temperature, entropy density, energy density, pressure and spin potential respectively. An important point to note is that the local thermodynamic relations do not generally hold in the quantum-statistical description of a relativistic fluid \cite{Becattini:2023ouz}, where the thermodynamic quantities, such as temperature, thermal velocity and spin potential can be unambiguously defined at the local thermodynamic equilibrium (LTE) \cite{Becattini:2014yxa}. In this work, we adhere to the traditional hydrodynamics viewpoint \cite{Bhattacharya:2011tra,Kovtun:2012rj} and assume that it is always possible to establish the local thermodynamic relations with a proper redefinition of the thermodynamic quantities in a state near equilibrium. Although the thermodynamic quantities defined in the two frameworks may share different values, they should approach the same values as the fluid evolves to the global thermodynamic equilibrium (GTE). %, indicating that the differences between the two approaches are at most of order $O\(\pd\)$ and should vanish in the GTE. We will not attempt to figure out these differences and simply assume that the local thermodynamic relations \eqref{Euler}-\eqref{firstlaw} are exact to the first order based on our present definition of the thermodynamic quantities, which we employ to describe the relativistic fluid. 
Additionally, we take a general antisymmetric spin potential $\o_{\m\n}$ without the requirement $\o_{\m\n}u^\n = 0$ even when it is conjugate to $\cR^{\m\n}$ in the local thermodynamic relations, as is case in the phenomenological formulation of spin hydrodynamics\cite{Hattori:2019lfp,Fukushima:2020ucl,Wang:2021ngp,Wang:2021wqq,Xie:2023gbo,Daher:2022xon}. One can also separate $\o_{\m\n}$ into rotation and boost parts as $\o_{\m\n}=r_{\m\n} + b_{\m\n}$ where
\begin{align}
r_{\m\n}=\e_{\m\n\r\s} r^\r u^\s ,\quad b_{\m\n}=2 b_{[\m} u_{\n]} \no
r^\s =  \fr{1}{2}\e^{\s\r\m\n}u_\r r_{\m\n},\quad b_\n = u^\m b_{\m\n} . \nonumber
\end{align}
%where $r_{\m\n}=\e_{\m\n\r\s} r^\r u^\s$, $b_{\m\n}=2u_{[\m} b_{\n]}$, $r^\s\equiv \fr{1}{2}\e^{\s\r\m\n}u_\r r_{\m\n}$ and $b_\m\equiv b_{\m\n}u^\n$. $\(\D_{\m}^{\;\r}u_{\n}u^{\s}-\D_{\n}^{\;\s}u_{\m}u^{\r}\)d\o_{\r\s}$ $b$-$\cR$ and $r$-$\cB$ We can understand the conjugations as follows.
The conjugations in \eqref{Euler}-\eqref{GibDuh} differ from the canonical formulations in \cite{Hongo:2021ona,Cao:2022aku} where the boost variables are absence and the spin potential is chosen as $r_{\m\n}$ with $r_{\m\n}u^\n = 0$. This difference is nontrivial. Although the Gibbs energy density $g$ from \eqref{Euler}, i.e.,
\begin{align}
g=\ep+p-T s=\fr{1}{2}\o_{\m\n}\cS^{\m\n}=\fr{1}{2}\(r_{\m\n}\cR^{\m\n}+b_{\m\n}\cB^{\m\n}\)=r_{\m}\cR^{\m}+b_{\m}\cB^{\m},
\end{align}
gets no contributions from $b_{\m\n}\cR^{\m\n}=r_{\m\n}\cB^{\m\n}=0$, % are identically zero, which may imply some reversible processes. 
the conjugations in \eqref{firstlaw} and \eqref{GibDuh}, 
\begin{align}
\fr{1}{2}\o_{\m\n}d\cS^{\m\n}=r_{\m}d\cR^{\m}+b_{\m}u_{\n}\e^{\m\n\r\s}\cR_\r d u_{\s}+b_{\m}d\cB^{\m}+\e_{\m\n\r\s}r^{\r}u^{\n}\cB^\n d u^{\m},\no
%r_{\m\n}d\cB^{\m\n}=r_{\m\n}\cB^\n d u^{\m},\no
%\cB^{\m\n}dr_{\m\n}=\cB^{\m\n}\e_{\m\n\r\s}r^{\r} d u^{\s},\no
\fr{1}{2}\cS^{\m\n}d\o_{\m\n}=\cR^{\m}dr_{\m}-b_{\m}u_{\n}\e^{\m\n\r\s}\cR_\r d u_{\s}+\cB^{\m}db_{\m}-\e_{\m\n\r\s}r^{\r}u^{\n}\cB^\n d u^{\m}, %\cR^{\m\n}db_{\m\n}=\cR^{\m\n}b_\m d u_{\n}, 
\end{align}
give $b$-$\cR$ and $r$-$\cB$ conjugations which are generally nonvanishing in the presence of vorticity where the non-inertial motion of fluid evolves along with the spin variables. %The corrections $r_{\m}d\cR^{\m}+b_{\m}d\cB^{\m}$ and $\cR^{\m}dr_{\m}+\cB^{\m}db_{\m}$ stem from the thermodynamic evolution of the spin variables within the domain orthogonal to four-velocity, while the crossed terms may characterize the non-inertial motion of fluid which accompanies the thermodynamic processes and drives the orthogonal domain to keep in step with the changing four-velocity. 
It may seem that the non-inertial $b$-$\cR$ and $r$-$\cB$ terms are not well defined contributions to thermodynamic potentials. Actually, the velocity dependence is just an artifact from the introducing of rotation $\cR^\m$ and boost $\cB^\m$ vectors orthogonal to four-velocity as thermodynamic quantities into the generalized local thermodynamic relations. % unless the fluid four-velocity or its direction is constant, i.e., $du=0$ or $du\pr u$. %the reference points and change rates scalar quantities, such as temperature, entropy density, energy density, pressure and so on, may be implied in their dependence on the non-inertial parameters
The general antisymmetric $\o_{\m\n}$ is more physically appropriate than $r_{\m\n}$ with $r_{\m\n}u^{\m}=0$, in the sense that $\o_{\m\n}$ can smoothly transition into the GTE to coincide with the general thermal vorticity $\vrp_{\m\n}\equiv\pd_{[\m}\b_{\n]}$ which may not necessarily be orthogonal to $u_{\m}$. % by definition not Moreover, in thermodynamics, the conjugation of the boost part $b_{\m\n}$ with $d\cR^{\m\n}$ is nonvanishing in \eqref{firstlaw} even if $b_{\m\n} \cR^{\m\n}=0$ in \eqref{Euler}. Therefore, it is nontrivial to adopt a general spin potential $\o_{\m\n}$ with both rotation and boost parts   %since if the spin potential is transverse to 
Especially, when the acceleration part $u^\m\vrp_{\m\n}$ is as strong as the spatial part $\fr{1}{2}\e^{\s\r\m\n}u_\r\vrp_{\m\n}$, we will see in the entropy-current analysis that the entropy principle with a general antisymmetric $\o_{\m\n}$ necessarily requires the presence of $\cB$ in the constitutive relations. %imposes significant constraints on . % satisfying with vanishing $b_{\m\n}$ part in $\o_{\m\n}$  %of the spin hydrodynamics. We will work on the  by  under such conditions. we choose  instead of 

We start with a general tensor decomposition in spin hydrodynamics as follows,
\begin{align}
&\Th^{\m\n}=\ep u^{\m} u^\n +p \D^{\m\n}+\mTh^{\m\n},\label{decpEm}\\
&\S^{\m\n\a}=\e^{\m\n\a\s}\(\cR_\s + \mS\, u_{\s}\) ,\label{decpAm}\\
%&J^{\m}=n\,u^{\m}+\mJ^{\m},\no
&s^{\m}=s\, u^{\m}+\ms^{\m},\label{decpEntr}
\end{align}
% the dual variable of spin density is defined as $\cR^\s\equiv\fr{1}{2}\e^{\s\a\m\n}u_\a\cR_{\m\n}$ and 
where $\D^{\m\n}\equiv\et^{\m\n}+u^\m u^\n$ is the transverse projection operator and the constitutive relations of the components with a circle are to be constrained by the entropy principle. 
To perform the entropy-current analysis, we derive the entropy production rate as follows. Taking the notations $\b\equiv 1/T$, $\b_\m\equiv \b u_\m$, $D\equiv u^{\n}\pd_{\n}$ and $\th\equiv \pd_{\n}u^{\n}$, we have
\begin{align}
\pd_\m s^{\m} = \pd_\m\(s\, u^{\m}+\ms^{\m}\)=D\, s + s\,\th + \pd_\m\ms^{\m} .
\end{align}
We replace the first term in the above expression using \eqref{firstlaw} to get
\begin{align}
\pd_\m s^{\m} = \b\[D \ep - \fr{1}{2}\o_{\m\n} D \(\cR^{\m\n} + \cB^{\m\n}\)\] + s\,\th + \pd_\m\ms^{\m} .
\end{align}
The two terms $D \ep$ and $-\fr{1}{2}\o_{\m\n} D \cR^{\m\n}$ in the square brackets can be substituted by the components $\mTh^{\m\n}$ and $\mS$ that are to be determined by the entropy principle. To proceed, we contract \eqref{Em} and \eqref{Am} with $u_\n$ and $\o_{\m\n}$ respectively to get
\begin{align}
D \ep & = - \(\ep + p \) \th + u_\n\pd_\m\mTh^{\m\n},\no 
-\fr{1}{2} \o_{\m\n} D \cR^{\m\n} & = \o_{\m\n}\[\fr{1}{2} \th \cR^{\m\n} + \pd_\a\(u^\n\cR^{\a\m}\)\]+\fr{1}{2}\e^{\a\m\n\s}\o_{\m\n}\pd_\a\(\mS u_\s\)+\o_{\m\n} \mTh^{[\m\n ]}. \label{EomTR}
\end{align}
We then obtain the entropy production rate as
\begin{align}
\pd_\m s^{\m} 
%=&\b\[-\(\ep +p\)\th+u_\n\pd_\m\mTh^{\m\n}\] + \b\[\o_{\m\n}\(\fr{1}{2} \th \cR^{\m\n} + \pd_\a\(u^\n\cR^{\a\m}\)\)+\fr{1}{2}\e^{\a\m\n\s}\o_{\m\n}\pd_\a\(\mS u_\s\)+\o_{\m\n} \mTh^{[\m\n ]}\]\no
%& - \b\fr{1}{2}\o_{\m\n} D \cB^{\m\n}+ s\,\th + \pd_\m\ms^{\m} \no
%=&\b\[T s-\(\ep +p\) + \fr{1}{2}\o_{\m\n} \(\cR^{\m\n} + \cB^{\m\n}\)\]\th + \pd_\m\ms^{\m} \no% - \b\fr{1}{2}\o_{\m\n} D \cB^{\m\n}
%& + \b\[u_\n\pd_\m\mTh^{\m\n} + \o_{\m\n} \mTh^{[\m\n ]}-\fr{1}{2} \o_{\m\n} \pd_\a\(u^{\a} \cB^{\m\n}\) + \o_{\m\n} \pd_\a\(u^\n\cR^{\a\m}\) + \fr{1}{2}\e^{\a\m\n\s}\o_{\m\n}\pd_{\a}\(\mS u_\s\)\] \no
%=&\b\[T s-\(\ep +p - \fr{1}{2}\o_{\m\n} \(\cR^{\m\n} + \cB^{\m\n}\)\)\]\th + \pd_{\m}\(\ms^{\m} + \mTh^{\m\n}\b_\n + \fr{1}{2}\e^{\m\a\n\s}\b\o_{\a\n}\mS u_\s\) \no
%& - \pd_\m\b_\n\mTh^{\m\n} + \b\o_{\m\n} \mTh^{[\m\n ]} - \fr{1}{2}\e^{\a\m\n\s}\pd_\a\(\b\o_{\m\n}\)\mS u_\s\no
%& + \b\o_{\m\n}\[\pd_\a\(u^\n\cR^{\a\m}\)-\fr{1}{2} \pd_\a\(u^{\a} \cB^{\m\n}\)\] \no
=&\[s-\b\(\ep +p - \fr{1}{2}\o_{\m\n} \(\cR^{\m\n} + \cB^{\m\n}\)\)\]\th + \pd_{\m}\(\ms^{\m} + \mTh^{\m\n}\b_\n + \fr{1}{2}\mS\,\e^{\m\a\n\s}\b\o_{\a\n} u_\s\) \no
& - \mTh^{(\m\n )} \pd_{\m}\b_{\n} - \mTh^{[\m\n ]}\(\pd_{\m}\b_{\n}-\b\o_{\m\n}\) - \fr{1}{2}\mS\,\e^{\a\m\n\s}\pd_\a\(\b\o_{\m\n}\) u_\s\no
& + \[\pd_\a\(u^\n\cR^{\a\m}\)-\fr{1}{2} \pd_\a\(u^{\a} \cB^{\m\n}\)\]\b\o_{\m\n}.\label{EntPro}
\end{align}
In the absence of $\cB$ and $b$, \eqref{EntPro} agrees with \cite{Hongo:2021ona,Cao:2022aku}, irrespective of the specific power counting scheme. %, while the  So far, the above result is exact without performing the gradient expansion. 
The entropy principle requires that \eqref{EntPro} is not only semipositive in general, but also zero in the GTE where thermal vorticity $\vrp_{\m\n}$ becomes a constant anti-symmetric tensor and
\begin{align}
\b_\m=c_\m+\vrp_{\m\n}x^\n,\quad \b\o_{\m\n}=\vrp_{\m\n}, \label{GTE}
\end{align}
with $c_\m$ being a constant four-vector. 

To explicitly seek the semipositivity of \eqref{EntPro} to the second order of gradient, we adopt a general power counting scheme where perturbation expansion of spin variables are independent of the conventional hydrodynamic variables and their gradients, % such as $u^\m$, %we adopt the power counting scheme% in \cite{Hattori:2019lfp,Fukushima:2020ucl,Wang:2021ngp,Wang:2021wqq,Daher:2022xon,Xie:2023gbo}, =\cR^{\m\n}+\cB^{\m\n}
\begin{align}
&\ep\sim p\sim \b \sim u^\m\sim O\(\pd^0\), \quad \cS^{\m\n}\sim O\(\d\) ,\no
&\o_{\m\n}\sim\vrp_{\m\n}\sim O\(\vrp\), \quad \vrp_{\m\n}-\b\o_{\m\n}\sim O\(\pd\), %,\quad \sim \vrs_{\m\n} \o_{\m\n}\sim\vrp_{\m\n} \sim O\(\o\)
\end{align}
%The differences in the gradient expansion will become apparent when assessing  in terms of quadratic forms that constrain the constitutive relations of $\mTh^{\m\n}$, $\ms^\m$ and $\mS$. $\vrs_{\m\n} \equiv \pd_{(\m}\b_{\n)}$ is the thermal shear tensor. W
where $O\(\d\)$ could be $O\(\hb\vrp\)$ if the spin susceptibility is $O\(\hb\pd^0\)$. %we have not specified the order of $\o_{\m\n}$ and $\vrp_{\m\n}$. 
In general, $O\(\vrp\)$ could range from $O\(\pd\)$ for hydrodynamics with an isotropy background to $O\(\pd^0\)$ for gyrohydrodynamics \cite{Cao:2022aku} with an anisotropic background. We count $O\(\d\)$ and $O\(\vrp\)$ independently of $O\(\pd\)$ so that the formulation of the spin hydrodynamics is applicable to a broad scale of the thermal vorticity instead of subject to a specific power counting scheme of it. %We assume that all the gradient terms except $\vrp_{\m\n}$ are of $O\(\pd\)$. 

%A general   
We aim to determine the constitutive relations of $\mTh^{\m\n}$, $\mS$ and $\ms^\m$ to $O\(\pd\)$ where \eqref{EntPro} should be semipositive to $O\(\pd^2\)$. In the precedent set by \cite{Hongo:2021ona}, the entropy production rate is ensured to be semipositive to $O\(\pd^2\)$ under a specific power counting scheme with $\cR^{\m\n}\sim\o_{\m\n}\sim O\(\pd\)$ and without the presence of $\cB^{\m\n}$, where the non-semipositive terms $\mTh^{[\m\n]}u_\n\(\pd_{\m}\b_{\s}-\b\o_{\m\s}\)u^\s$ and $\b\o_{\m\n}\pd_\a\(u^\n\cR^{\a\m}\)$ in \eqref{EntPro} can be neglected as $O\(\pd^3\)$. However, in a broad scale of the thermal vorticity, these non-semipositive terms are of $O\(\pd\vrp\d\)$ which are generally non-ignorable and therefore have to be cancelled out. % by the entropy production rate from $\mTh^{\m\n}$ and $\mS$.  $O\(\pd^2\d\)$ and
By noting \eqref{Euler}, we drop the first term in the first line of \eqref{EntPro}. Taking the GTE limit \eqref{GTE} in \eqref{EntPro}, one has
\begin{align}
0=\pd_\m s_{\mathrm{GTE}}^{\m}=& \pd_{\m}\(\ms_{\mathrm{GTE}}^{\m} + \mTh_{\mathrm{GTE}}^{\m\n}\b_\n + \fr{1}{2}\mS_{\mathrm{GTE}}\,\e^{\m\a\n\s}\vrp_{\a\n} u_\s\) \no
& + \[\pd_\a\(u^\n\cR^{\a\m}\)-\fr{1}{2} \pd_\a\(u^{\a} \cB^{\m\n}\)\]\vrp_{\m\n}.\label{EntProGTE}
\end{align}
Noting the nonvanishing terms in the last brackets of \eqref{EntProGTE}, it is evident that $\ms^{\m} \neq -\mTh^{\m\n}\b_\n - \fr{1}{2}\mS\,\e^{\m\a\n\s}\vrp_{\a\n} u_\s$ in general. Therefore, terms involving $\cR$ and $\cB$ must be present in either $\mTh^{\m\n}$, $\ms^\m$ or $\mS$ to offset the above nonvanishing terms. 
Note that, in the vicinity of GTE, these nonvanishing terms arise from the leading-order term $\e^{\m\n\a\s}\cR_\s$ in the spin current \eqref{decpAm}. As pointed out in \cite{Hattori:2019lfp}, the entropy production rate from the leading-order of the spin current is zero if spin and orbital angular momentum are separately conserved\footnote{Although there are no additional symmetries beyond the Lorentz group that would allow for individual conservation laws in field theory, and ideal spin hydrodynamics does not exist, we can establish an ad hoc criterion for formulating spin hydrodynamics, which suggests that a hydrodynamic framework should be non-dissipative at the leading-order in the conservation limit of the currents involved.}. At the lowest order of the non-conservation equation \eqref{Am} of the spin current, the dissipation of spin only stem from the source/absorption term $\mTh^{[\m\n ]}$. %We assume that the non-dissipative nature of the leading-order spin current holds ture  %is means that the leading-order spin current is non-dissipative while the  currents in the constitutive relations, similar to the case of the background $\ep u^{\m} u^\n +p \D^{\m\n}$ in \eqref{decpEm}. Thus, without leaving any semipositive contribution. This condition may not be physically realistic for a real fluid, as 

We separate the non-dissipative parts from the dissipative parts by marking the former with subscript $\d$ and the latter with tick, i.e., $\mTh^{\m\n}=\Th_{\d}^{\m\n}+\cTh^{\m\n}$, $\mS=\S_{\d}+\chS$ and $\ms^\m=s_{\d}^\m+\chs^\m$. % involved in eliminating these non-semipositive terms irrelevant  where $\Th_{\d}^{\m\n}$, $\S_{\d}$ and $s_{\d}^\m$ are vanishing when both $\cR_{\m\n}$ and $\cB_{\m\n}$ are zero, while $\cTh^{\m\n}$, $\chS$ and $\chs^\m$ are $\d$ independent. 
At this stage, we manifest the assumption that the constitutive relations of $\mTh^{\m\n}$, $\ms^\m$ and $\mS$, as expressions in terms of the spin hydrodynamic variables $\b$, $u^\m$, $\o^{\m\n}$ and $\cS^{\m\n}$, consistently satisfy the entropy principle, i.e., %  in $\Th_{\d}^{\m\n}$, $ s_{\d}^\m$ and $\S_{\d}$
\begin{align}
\exists \, \mTh^{\m\n},\ms^{\m},\mS \text{ as functions of }\b,u^{\m}, \o^{\m\n}\text{ and }\cS^{\m\n} \; \forall \, \b,u^{\m}, \o^{\m\n}\text{ and }\cS^{\m\n}:\pd_\m s^\m\ge 0 .\label{cnstr}
\end{align}
%$a^\m$ is an arbitrary vector with $a\cdot u=0$ representing the possible anisotropy in the background due to a strong thermal vorticity. 
Here the $\cS^{\m\n}$ dependent parts of the constitutive relations are to cancel out the non-semipositive terms in \eqref{EntPro} where we take $\cS^{\m\n}$ as extra free variables besides $\b,u^{\m}$ and $\o^{\m\n}$\footnote{Note that $\cS^{\m\n}$ is free in the sense that the the dependence of $\cS^{\m\n}$ on $\b,u^{\m}$ and $\o^{\m\n}$ could vary with specific type of fluid and physical regime while the constitutive relations should satisfy the entropy principle in general.}. %since the dependence could vary with different while constituent particles through out of the spin hydrodynamic regime . One does not have to worry about the dependence of
The entropy production rate is written as
\begin{align}
\pd_\m s^{\m} & = \pd_\m\[\chs^{\m} - \cTh^{\m\n}\b_{\n} + \fr{1}{2}\chS \e^{\m\a\n\s}\b\o_{\a\n} u_\s\]\label{EntProIr}\\
&-\cTh^{(\m\n)}\pd_\m\b_{\n} - \cTh^{[\m\n]}\(\pd_{\m}\b_{\n}-\b\o_{\m\n}\) - \fr{1}{2}\chS \e^{\a\m\n\s}\pd_\a\(\b\o_{\m\n}\) u_\s\no
& + \pd_\m  s_{\d}^\m + \pd_\m\Th_{\d}^{\m\n}\b_\n +\[\Th_{\d}^{\m\n} + \pd_\a\(u^\n\cR^{\a\m}\) - \fr{1}{2} \pd_\a\(u^{\a} \cB^{\m\n}\) + \fr{1}{2} \e^{\a\m\n\s} \pd_\a\(\S_{\d}\, u_\s\)\] \b\o_{\m\n}. \nonumber
\end{align}
The dissipative part $\cTh^{\m\n}$ can be decomposed into the irreducible tensor basis \cite{Hattori:2019lfp,Fukushima:2020ucl,Daher:2022xon}(see also \cite{Baier:2007ix,Denicol:2012cn,Molnar:2013lta}) as follows, %  separated , $\chS$ and $\chs^\m$
\begin{align}
\cTh^{(\m\n)}=2u^{(\m}h^{\n)} + \t^{\m\n},\qquad \cTh^{[\m\n]}=2u^{[\m} q^{\n]} + \ph^{\m\n}\label{dcpDis},
\end{align}
where the dissipative currents satisfy $\t^{\m\n}=\t^{\n\m},\,\ph^{\m\n}=-\ph^{\n\m},\,u_\m h^\m=u_\m q^\m=u_\m\t^{\m\n}=u_\m\ph^{\m\n} = 0$. We have $u_\m\cTh^{\m\n}u_\n=\ep$ while $u_\m\Th_{\d}^{\m\n}u_\n$ is not necessarily zero. Moreover, in contrast to \cite{Biswas:2023qsw}, we do not require $u_\m\ms^\m\le 0$. This is because we assume that the local thermodynamic relations \eqref{Euler}-\eqref{firstlaw} hold near LTE, where all the thermodynamic variables, including the entropy density $s$, are extended to be applicable out of equilibrium. Therefore, a general entropy density $s$ evolving towards equilibrium is not identically equal to the maximum value that is to be reached in equilibrium. %  guaranteed to be state

In addition to the dissipative parts in the entropy production rate, we have collected all the non-dissipative components into the last line of \eqref{EntProIr}. Explicitly, the entropy principle requires that the sum of the non-dissipative terms gives zero entropy production rate
\begin{align}
&\pd_\m  s_{\d}^\m + \pd_\m\Th_{\d}^{\m\n}\b_\n \no
&\quad + \[\Th_{\d}^{\m\n} + \pd_\a\(u^\n\cR^{\a\m}\) - \fr{1}{2} \pd_\a\(u^{\a} \cB^{\m\n}\) + \fr{1}{2} \e^{\a\m\n\s} \pd_\a\(\S_{\d}\, u_\s\)\] \b\o_{\m\n}=0. \label{nonDis}
\end{align}
We consider the non-dissipative constitutive relations of $\Th_{\d}^{\m\n}$, $ s_{\d}^\m$ and $\S_{\d}$ to all orders as solutions to \eqref{nonDis}. %without any semipositive contribution \footnote{We have excluded $O\(\d^2\)$ terms in the constitutive relations of $\Th_{\d}^{\m\n}$, $ s_{\d}^\m$ and $\S_{\d}$ which may absorb the $O\(\pd\vrp\d\)$ background as cross terms into quadratic forms since we expect no extra dissipative constitutive relations from the background terms.}. 
% we categorize the terms in \eqref{nonDis} into specific groups of $O\(\pd^0\vrp^{1|2}\d\)$, $O\(\pd\vrp^{0|1}\d\)$, $O\(\pd^2\vrp^{0}\d\)$ and higher orders. %In general, one may also consider the $O\(\pd^2\vrp^0\d^2\)$ and $O\(\pd^0\vrp^2\d^2\)$ terms, which, together with the present terms, yield quadratic forms such as $\(\fr{1}{\sqrt{c}}\pd\d + \sqrt{c}\,\o\)^2$ and $\(\fr{1}{\sqrt{c}}\vrp\d+\sqrt{c}\,\pd u\)^2$ to absorb the cross terms $\pd\(u \d\) \o$, where $c$ is the positive kinetic coefficient. However, such terms, if allowed, lead to components like $\fr{1}{c}\pd\d^2$ and $\fr{1}{c}\vrp\d^2$ in the constitutive relations, which are not only %In this situation, the non-semipositive terms in \eqref{EntProIr} have to be non-dissipative, and therefore should be canceled out instead of being absorbed into quadratic forms.  % in \eqref{nonDis} from the parts involving of order $O\(\pd^0\o^0\cR^0\cB^0\)$ and composed of
%At the first order of gradient, Accordingly, 
For this purpose, we explicitly write $\Th_{\d}^{\m\n}$ and $ s_{\d}^\m$ into the terms of $O\(\pd^0\o^{0}\d\)$, $O\(\pd^0\o\d\)$, $O\(\pd\o^{0}\d\)$ and higher orders in a general form as 
\begin{align}
&\Th_{\d}^{\a\s} = \Th_{0}^{\a\s} + \Th_{\o}^{\a\s\m\n} \o_{\m\n} + \Th_{\pd}^{\a\s} - \fr{1}{2} \e^{\m\a\s\n} \pd_\m\(\S_{\d}\, u_\n\) + O\(\pd\o\d\) ,\no % \Th_{\o\pd}^{\a\s\m\n}\o_{\m\n} + \Th_{\pd\o}^{\a\s\r\m\n}\pd_\r\o_{\m\n}
& s_{\d}^\a = s_{0}^{\a} + s_{\o}^{\a\m\n}\b\o_{\m\n} + s_\pd^\a + O\(\pd\o\d\) ,\label{ConDcp}% s_{\o\pd}^{\a\m\n}\o_{\m\n} + s_{\pd\o}^{\a\r\m\n}\pd_\r\o_{\m\n} \quad \S_\d=\S_0 + \S_\o^{\m\n}\o_{\m\n} 
\end{align}
where the $O\(\pd^0\o^0\d\)$ components $\Th_{0,\o},s_{0,\o}$ and $O\(\pd\o^0\d\)$ components $\Th_{\pd},s_{\pd}$ are expressions in terms of $\b,u^\m$ and $\cS^{\m\n}$. %The $O\(\pd\o^0\d\)$ parts in $\S_{\d}$ only contribute $O\(\pd^2\o\d\)$ terms in \eqref{EntProIr} and therefore are not relevant to us. The $O\(\pd^0\o^0\d\)$ terms in $\S_{\d}$ only yield $O\(\pd\o\d\)$ terms in \eqref{EntProIr}.  
Now we collect the terms involving $\b\o_{\m\n}$, $\pd_\a\(\b\o_{\m\n}\)$ and $\b\o_{\a\s} \b\o_{\m\n}$ into
\begin{align}
X^{\m\n}\b\o_{\m\n} + Y^{\a\m\n}\pd_\a\(\b\o_{\m\n}\) + \Th_{\o}^{\a\s\m\n}T \b\o_{\a\s} \b\o_{\m\n} ,\label{EntProOmg} %+ \pd_\a s_{\pd}^\a + \pd_\a\Th_{\pd}^{\a\s}\b_\s
\end{align}
where $X^{\m\n}$ and $Y^{\a\m\n}$ are defined as %the $O\(\pd\o\d\)$ terms are gathered in two independent parts involving $\b\o_{\m\n}$ and $\pd_\a\(\b\o_{\m\n}\)$ denoted as 
\begin{align}
&X^{\m\n}\equiv\pd_\a s_{\o}^{\a\m\n} + \pd_\a\(\Th_{\o}^{\a\s\m\n}T\)\b_\s + \Th_{\pd}^{\m\n} + \pd_\a\(u^\n\cR^{\a\m} - \fr{1}{2} u^{\a} \cB^{\m\n}\) ,\no
&Y^{\a\m\n}\equiv s_{\o}^{\a\m\n} + \Th_{\o}^{\a\s\m\n} u_\s.
\end{align}
%, the first line and the first square brackets in the second line. 
The two parts must vanish for any values of $\b\o_{\m\n}$ and $\pd_\a\(\b\o_{\m\n}\)$, resulting in the constraints 
\begin{align}
X^{[\m\n]}=0 \;\text{ and }\; Y^{\a[\m\n]}=0. \label{cnstrXY}
\end{align} %Similar default preconditions apply to the remainder of this work. 
%where the implicit precondition $\exists\{\Th_\pd,\Th_\o,s_\o\}\(\b,u,\o,\d\) \forall \{\b,u,\o,\d\}$ is omitted for the sake of concise notation. 
Likewise, to ensure the term $\Th_{\o}^{\a\s\m\n}\b\o_{\a\s} \b\o_{\m\n}$ vanishes for arbitrary values of $\b\o_{\a\s} \b\o_{\m\n}$ in \eqref{EntProOmg}, $\Th_{\o}^{\a\s\m\n}$ can run through several switches as follows
\begin{align}
%&\exists\Th_\o\(\b,u,\o,\d\) \forall \{\b,u,\o,\d\}: \no & \myquad[9] 
\Th_{\o}^{\a\s[\m\n]}=0 \;\text{ or }\; \Th_{\o}^{[\a\s]\m\n}=0 \;\text{ or }\; \Th_{\o}^{\a\s\m\n}=-\Th_{\o}^{\m\n\a\s} \;\text{ or }\; \Th_{\o}^{\a\s\m\n}=-\Th_{\o}^{\n\m\s\a} . \label{cnstrTho}
\end{align} 
We then combine \eqref{cnstrXY} and \eqref{cnstrTho} to constrain the solutions to \eqref{nonDis}. As a straightforward application of these constraints, one can readily confirm that 
\begin{align}
\Th_{\pd}^{\m\n}=0 \;\text{ and }\; \Th_{\o}^{\a\s\m\n}=0 \;\text{ and }\; \eqref{cnstrXY} \;\to\; s_{\o}^{\a[\m\n]}=0 \;\to\; \pd_\a\(\cR^{\a[\m}u^{\n]} - \fr{1}{2} u^{\a} \cB^{\m\n}\)=0,
\end{align}
leading to a contradiction as the left-hand side of the final equation is not identically zero. This implies that the stress-energy tensor must depend on $S^{\m\n}$ at $O\(\pd\d\)$ or $O\(\o\d\)$.
In general, one can analyze the terms to all orders in \eqref{nonDis} to obtain a complete constraint for the solution. Nevertheless, given that \eqref{nonDis} must hold order by order, we concentrate exclusively on the constraints related to the $O\(\pd\d\)$ and $O\(\o\d\)$ terms. It will become apparent in the next section that the components dependent on $\o$ within these orders are sufficient to illustrate the difficulties in upholding the entropy principle in the absence of boost variables. %The $O\(\pd^2\o^0\d\)$ part involving double $\pd$ operators is cumbersome to analyse in groups, we leave it for future study. 
\iffalse The constraint \eqref{cnstr} is concretized to
\begin{align}
\(\exists\{\Th_\pd,s_\pd\}\(\b,u,\o,\d\) \forall \{\b,u,\o,\d\}:\pd_\m s_\pd^\m + \pd_\m\Th_\pd^{\m\n}\b_\n = 0\) \;\text{ and }\; \no
\exists\{\Th_\pd,s_\pd\}\(\b,u,\o,\d\) \forall \{\b,u,\o,\d\}:\pd_\m s_\pd^\m + \pd_\m\Th_\pd^{\m\n}\b_\n = 0. \label{cnstr1}
\end{align}
\fi

\section{entropy principle in the absence of boost variables}\label{SecIss}

%in this section, In the next section, to illustrate that the independent modes associated with the boost symmetry are crucial within a specific power counting scheme \footnote{Note that the constitutive relations in spin hydrodynamics may vary with the specific power counting scheme, given that the entropy-current analysis involves derivative expansion.} to uphold the entropy principle. 
%semiquantitative for boost coupling
%qualitative for non-semipositvie terms
%and that spin polarization represents an additional independent dynamical mode within the realm of spin hydrodynamics. 
%of spin density attached to the spatial rotation symmetry there are only three independent components in the spin density 
We now investigate the framework in the abscence of the degree of freedom related to the boost symmetry where there are only seven independent dynamical variables with four from relativistic translation symmetry and three from rotation symmetry. In such circumstances, it is necessary to select three out of the ten equations in \eqref{Em}-\eqref{Am} as redundant in order to avoid overdetermination. As pointed out in \cite{Hongo:2021ona}, the physically meaningful choice is to consider the three equations ensuing from the boost symmetry in \eqref{Am} as redundant identities since the boost variables are vanishing. The identities in the local rest frame are obtained by setting $\m=0,\n=i$ or $\m=i,\n=0$ in \eqref{Am}, while the covariant form is manifested by projecting \eqref{Am} onto $u_\n$ as
\begin{align}
&\(\pd_\a\S^{\a\m\n}+2\Th^{[\m\n]}\)u_\n=0\no
&\qquad\Rightarrow\quad \fr{1}{2} \e^{\a\m\n\s} u_\n\pd_\a\(\cR_\s + \S_{\d} u_\s\) + \Th_{\d}^{[\m\n]}u_\n + \fr{1}{2} \e^{\a\m\n\s} u_\n\chS\,\pd_\a u_\s + q^\m=0.
\end{align}
Noting that $q^\m$ at $O\(\pd\d\)$ and $\chS$ at $O\(\d\)$ are both zero to ensure the semipositivity of the dissipative parts in \eqref{EntProIr}, we isolate the $O\(\pd\d\)$ terms from the other parts in the above identity, 
\begin{align}
q^\m & = - \fr{1}{2} \e^{\a\m\n\s} u_\n\chS\,\pd_\a u_\s,\label{idQ}\\
\Th_{\d}^{[\m\n]}u_\n & = - \fr{1}{2} \e^{\a\m\n\s} u_\n\pd_\a\(\cR_\s + \S_{\d} u_\s\), \label{idTh} 
\end{align}
%where we have taken $\S_{\d}=0$ as previously argued. 
where the identity at $O\(\pd\d\)$ in \eqref{idTh} should hold for arbitrary $\cR$. Utilizing the identities \eqref{idQ}-\eqref{idTh} as the result of the vanishing boost variables, we can demonstrate that it is not possible to cancel out the non-semipositive term $\pd_\a\(u^\n\cR^{\a\m}\)\b\o_{\m\n}$ in \eqref{nonDis}. %Since the cancellation of this term only involves the $\cR$ dependent terms, without loss of generality, we can analyse the $\cR$ dependent parts separately from the dissipative parts, where the former are represented by the last line of \eqref{EntProIr} while the latter are reflected in the first two lines. 

For the $\cR$ dependent parts, we further collect the $O\(\pd\o^0\d\)$ and $O\(\pd^0\o\d\)$ terms in \eqref{idTh} to obtain the extra constraints from the vanshing of $\cB$ as 
\begin{align}
& \Th_{\pd}^{[\m\n]}u_\n = - \fr{1}{2} \e^{\a\m\n\s} u_\n\pd_\a\cR_\s,\label{idThp}\\%\pd_\m\cR_\n
& \Th_{\o}^{[\m\n]\a\s}u_\n\o_{\a\s}=0 \quad\to\quad \Th_{\o}^{[\m\n][\a\s]}u_\n = 0 . \label{idTho}
\end{align} 
%Removing the $O\(\pd^2\o^0\d\)$ parts presents a challenge. For simplicity,  we will focus on the $\o$ dependent parts. 
We now examine the combined constraints on $\Th_{\o}$ in \eqref{cnstrXY}-\eqref{cnstrTho} and \eqref{idThp}-\eqref{idTho}. For the first switch in \eqref{cnstrTho}, one has 
\begin{align}
&\Th_{\o}^{\a\s[\m\n]}=0 \;\text{ and }\; Y^{\a[\m\n]}=0 \,\to\, s_{\o}^{\a[\m\n]}=0 \;\to\; X^{[\m\n]}= \Th_{\pd}^{[\m\n]} + \pd_\a\( \cR^{\a[\m}u^{\n]}\)=0 , 
\end{align}
which obviously contradicts the identity \eqref{idThp}. Thus, we get $\Th_{\o}^{\a\s[\m\n]}\neq 0$. 
\iffalse
Then, the rest three switches in \eqref{cnstrTho} combining with \eqref{idTho} becomes %
\begin{align}
\(\Th_{\o}^{[\a\s]\m\n}=0 \;\text{ or }\; \Th_{\o}^{\a\s\m\n}=-\Th_{\o}^{\m\n\a\s} \;\text{ or }\; \Th_{\o}^{\a\s\m\n}=-\Th_{\o}^{\n\m\s\a}\) \;\text{ and }\; \Th_{\o}^{[\a\s][\m\n]}u_\s = 0. \label{cnstrTho1}
\end{align} 
\fi

For the rest three switches in \eqref{cnstrTho}, using \eqref{idThp} in \eqref{cnstrXY} while noting $u\cdot u=-1$ and $u\cdot\cR=0$, we get
\begin{align}
%&X^{[\m\n]} = \pd_\a s_{\o}^{\a[\m\n]} + \pd_\a\(\Th_{\o}^{\a\s[\m\n]}T\)\b_\s + \Th_{\pd}^{[\m\n]} + \pd_\a\(\cR^{\a[\m} u^{\n]}\) ,\no
%&Y^{\a[\m\n]} =  s_{\o}^{\a[\m\n]} + \Th_{\o}^{\a\s[\m\n]} u_\s. \no
%&X^{[\m\n]} = - \Th_{\o}^{\a\s[\m\n]}\(T\pd_\a\b u_\s + \pd_\a u_\s\) + \Th_{\pd}^{[\m\n]} + \pd_\a\(\cR^{\a[\m} u^{\n]}\) ,\no
%& \Th_{\pd}^{[\a\s]}u_\s = - \fr{1}{2} \e^{\m\a\s\n} u_\s\pd_\m\cR_\n\no
&\Th_{\o}^{\a\s[\m\n]}u_\n u_\s\pd_\a\ln\b + \Th_{\o}^{\a\s[\m\n]}u_\n\pd_\a u_\s - W_1^{\a\m\n}\cR_\n\pd_\a\(u\cdot u\) - W_2^{\a\m}\pd_\a\(u\cdot\cR\) \no
&= \pd_\a\cR^{\a[\m}u^{\n]}u_\n - \fr{1}{2} \e^{\a\m\n\s} u_\n\pd_\a\cR_\s
= \fr{1}{2}\pd_\a\(u^\a\cR^{\m\n}\)u_\n 
%=-\fr{1}{2}\pd_\a\(u^\a\e^{\m\n\s\l}u_\s\cR_\l\)u_\n\no 
=-\fr{1}{2} u^\a\e^{\m\n\s\l}\cR_\l u_\n\pd_\a u_\s , 
%=-\fr{1}{2}\pd_\a\(T u^\a\e^{\m\n\s\l}\b_\s\cR_\l\)u_\n ,
\end{align}
%$\e^{\m\n\a\s}\cR_\s = u^{\m} \cR^{\n\a} + u^\n\cR^{\a\m} + u^\a\cR^{\m\n} $
where $W_1$ and $W_2$ could be any dimensionless tensors as expressions in terms of $\b,u^\m$ and $\cR^{\m}$. We have used the first identity of \eqref{IDcS} in the second equality. Given that the above equation holds for arbitrary $\pd_\a\ln\b$, $\pd_\a\cR_\s$ and $\pd_\a u_\s$, we have the constraints 
\begin{align}
&\Th_{\o}^{\a\s[\m\n]}u_\n u_\s=0 \;\text{ and }\; %\bigg(\exists \{W_1,W_2\}\(u,\b\)\forall u,\b:  \no &\myquad[10]
W_2^{\a\m}u^\s=0 \no 
&\;\text{ and }\; \Th_{\o}^{\a\s[\m\n]}u_\n -W_1^{\a\m\n}\cR_\n u^\s-W_2^{\a\m}\cR^\s-W_3^{\a\m}a^\s=-\fr{1}{2} u^\a\e^{\m\n\s\l}\cR_\l u_\n ,\label{idThpo}
\end{align}
which renders 
\begin{align}
&0=\Th_{\o}^{\a\s[\m\n]}u_\n u_\s=-W_1^{\a\m\n}\cR_\n \;\to\; W_1^{\a\m\n} =0 \no
&\text{ and }\;W_2^{\a\m}u^\s=0 \;\to\; W_2^{\a\m}=0 \;\to\; \Th_{\o}^{\a\s[\m\n]}u_\n=-\fr{1}{2} u^\a\e^{\m\n\s\l}\cR_\l u_\n\no
&\to \Th_{\o}^{[\a\s][\m\n]}u_\n=-\fr{1}{2} u^{[\a} \e^{\s] \m\n\l}\cR_\l u_\n\neq 0. \label{cnstrTho2}
\end{align}
This excludes the second switch in \eqref{cnstrTho}, i.e., $\Th_{\o}^{[\a\s]\m\n}\neq0$. 

The last two switches in \eqref{cnstrTho} combined with \eqref{idTho} reduce to 
\begin{align}
&\(\Th_{\o}^{\a\s\m\n}=-\Th_{\o}^{\m\n\a\s} \;\text{ or }\; \Th_{\o}^{\a\s\m\n}=-\Th_{\o}^{\n\m\s\a}\) \;\text{ and }\; \Th_{\o}^{[\a\s][\m\n]}u_\s = 0 \no
&\,\to\,\(\Th_{\o}^{\a\s[\m\n]}=-\Th_{\o}^{[\m\n]\a\s} \;\text{ or }\; \Th_{\o}^{\a\s[\m\n]}=-\Th_{\o}^{[\n\m]\s\a}\) \;\text{ and }\; \Th_{\o}^{[\a\s][\m\n]}u_\s = 0 \no
& \,\to\,\Th_{\o}^{[\a\s][\m\n]}u_\n=-\Th_{\o}^{[\m\n][\a\s]}u_\n =0 \;\text{ or }\; \Th_{\o}^{[\a\s][\m\n]}u_\n=-\Th_{\o}^{[\n\m][\s\a]}u_\n =0 , %\Th_{\o}^{\a\s\m\n}=-\Th_{\o}^{\m\n\a\s} \,\to\, 
\end{align}
which is also ruled out by \eqref{cnstrTho2}. 
Hence, there is no consistent result for $\Th_{\o}$ to ensure the vanishing of the $O\(\pd\o\d\)$ and $O\(\pd^0\o^2\d\)$ parts in \eqref{nonDis}. In other words, with a general antisymmetric spin potential $\o_{\m\n}$ and vanishing boost variables $\cB^{\m\n}$, it is generally not possible for the constitutive relations of spin hydrodynamics to satisfy the entropy principle. Note that \eqref{cnstrXY}-\eqref{cnstrTho} and \eqref{idTho} are constraints resulting from a general antisymmetric spin potential $\o_{\m\n}$. A meaningful complement would be to apply the entropy-current analysis presented in this study to the case of a special spin potential $r_{\m\n}$ with $r_{\m\n}u^\n=0$, as utilized in \cite{Hongo:2021ona,Cao:2022aku}, to investigate the existance of a solution for $\Th_{\d}$ and $s_{\d}$. We will not attempt to address this issue here. %find out possible results consistent with the entropy principle in our present power counting scheme. 

\section{Consistent First-Order Spin hydrodynamics}\label{SecConstnt}

The challenge in adhering to the entropy principle arises from the lack of degree of freedom associated with boost symmetry.  Therefore, it is inevitable to activate the boost variables so that the canonical formulation of spin hydrodynamics aligns with the entropy principle. In this scenario, the boost components of the conservation law \eqref{Am} are independent equations, rather than being fixed as identities like in \eqref{idQ}-\eqref{idTh}. %Consequently, the dissipative constitutive relation of $q^\m$ is constrained as in \eqref{ConRel}. 
The semipositivity of the dissipative parts is ensured by adopting the constitutive relations that are basically the same as those in \cite{Hattori:2019lfp,Fukushima:2020ucl,Daher:2022xon}, 
\begin{align}
h^\m &= -T h^{\m\n\a} \pd_\a\b_\n ,%= -T\k \D^{\m\n}\(D \b_\n-\pd_\n\b\),\no%\(\pd_\a\b_\n+\pd_\n\b_\a\)
&q^\m  &= -T q^{\m\n\a} \(\pd_\a\b_\n - \b\o_{\a\n}\) ,\no %&=-T\k_s \D^{\m\n}u^{\a}\(\pd_\a\b_\n - \pd_\n\b_\a - 2\b\o_{\a\n}\)\no%= -T\k_s \D^{\m\n}\(D \b_\n + \pd_\n\b\),
\t^{\m\n}&=-T\t^{\m\n\a\s}\pd_\a\b_{\s},
&\ph^{\m\n}&=-T\ph^{\m\n\a\s}\(\pd_\a\b_{\s}-\b\o_{\a\s}\) ,\label{ConRel}\\
\chS &= - \fr{1}{2}T \x \e^{\m\n\a\s}u_\s\pd_\a\(\b\o_{\m\n}\),
&\chs^{\m} &= \b h^{\m} - \b q^{\m} - \fr{1}{2}\chS \e^{\m\a\n\s}\b\o_{\a\n} u_\s ,\nonumber
\end{align}
where 
\begin{align}
h^{\m\n\a} &\equiv \k \D^{\m(\n}u^{\a)},\myquad[4]
q^{\m\n\a} \equiv \k_s \D^{\m[\n}u^{\a]},\no
\t^{\m\n\a\b} &\equiv 2\et\[\fr{1}{2}\(\D^{\m\a}\D^{\n\b}+\D^{\m\b}\D^{\m\n}\)-\fr{1}{3}\D^{\m\n}\D^{\a\b}\]+\z\D^{\m\n}\D^{\a\b} ,\no
\ph^{\m\n\a\b} &\equiv \fr{1}{2}\et_s\(\D^{\m\a}\D^{\n\b}-\D^{\m\b}\D^{\n\a}\) , 
\end{align}
with positive coefficients $\k,\,\k_s,\,\et,\,\z,\,\et_s$ and $\x$. In the case $O\(\vrp\) \sim O\(\pd^0\)$, the dissipative currents in \eqref{ConRel} can be further decomposed according to the anisotropy in gyrohydrodynamics \cite{Cao:2022aku}. %Although we do not show the explicit anisotropic decomposition, the entropy-current analysis performed in this work also applies to it since we only case about the symmetry of indices. 

As regards the non-dissipative terms in \eqref{nonDis}, it is known that there is a solution corresponding to a pseudogauge transformation from the phenomenological formulation of spin hydrodynamics\cite{Daher:2022xon},
\iffalse
Generally, an antisymmetric tensor $\cT^{\m\n}$ can be decomposed as
\begin{align}
&\cT^{\m\n}=\e^{\m\n\a\s} \cA_\a u_\s + 2u^{[\m}\cV^{\n]},\nonumber
\end{align}
where 
\begin{align}
&\cA_\a=\fr{1}{2}\e_{\a\m\n\s}\cT^{\m\n}u^\s,\qquad \cV^\m=\cT^{\m\n}u_\n\nonumber
\end{align}
with $\cA\cdot u=\cV\cdot u=0$ and $\cA,\,\cV$ being an axial vector and a polar vector respectively. We then 
Noting the identities $\e^{\m\n\a\s}\cR_\s = u^{\m} \cR^{\n\a} + u^\n\cR^{\a\m} + u^\a\cR^{\m\n} $ and $0 = u^{\m} \cB^{\n\a} + u^\n\cB^{\a\m} + u^\a\cB^{\m\n} $, the solution then can be explicitly written as
\fi
\begin{align}
\Th_{\d}^{\m\n} = -\pd_\a\(\cR^{\a\m}u^{\n}+\cB^{\a\m}u^{\n}\),\myquad[2] s_{\d}^\m = 0,\myquad[2] \S_{\d} = 0 .\label{SolBst}
\end{align}
%which eliminates all the $\d$ dependent parts and gives a canonical formulation consistent with entropy principle. 
Actually, it has been point out in \cite{Hattori:2019lfp} that given a formulation of spin hydrodynamics with $(\Th,\S)$ that satisfies entropy principle, a pseudogauge transformation always renders another consistent formulation $(\Th',\S')$ since the entropy production rate in the entropy-current analysis remains unchanged, though different pairs $(\Th,\S)$ and $(\Th',\S')$ are generally thermodynamically inequivalent \cite{Becattini:2011ev,Becattini:2012pp}.
Therefore, a general pseudogauge-transforming solution is
\begin{align}
\Th_{\d}^{\m\n} = -\pd_\a\(\cR^{\a\m}u^{\n}+\cB^{\a\m}u^{\n}\) - \fr{1}{2} \e^{\a\m\n\s} \pd_\a\(\S_{\d}\, u_\s\),\myquad[2] s_{\d}^\m = 0,\myquad[2] \S_{\d}\(\b,u,\o,\cS\) ,\label{SolBstGen}
\end{align}
where $\S_{\d}$ can be any possible scalar expression in terms of $\b,u,\o$ and $\cS$ with $\S_{\d}(\b,u,\o=0,\cS=0)=0$ so that $\S_{\d}$ vanishes in spinless limit. 
In addition, the entropy-gauge transformation \cite{Becattini:2023ouz} $s_\d^\m=\pd_\a A^{\a\m}$ with $A^{\a\m}=-A^{\m\a}$ gives an extra general solution where $A^{\a\m}$ could be any possible antisymmetric tensor expression in terms of $\b,u,\o$ and $\cS$. %, e.g., $A^{\a\m}=\sum_{\d=\cR,\cB}C_{\d} \b\d^{\a\m}$ with $C_{\d}$ being dimensionless constants. 
It remains to be seen whether there are non-dissipative solutions $\Th'_{\d}$, $s'_{\d}$ and $\S'_{\d}$ beyond the pseudogauge and the entropy-gauge transformations.
Concretely, such solutions are constrained by the entropy principle as
\begin{align}
\pd_\m  {s'}_{\d}^{\m} + \pd_\m{\Th'}_{\d}^{\m\n}\b_\n + \[{\Th'}_{\d}^{\m\n} + \fr{1}{2} \e^{\a\m\n\s} \pd_\a\(\S'_{\d}\, u_\s\)\] \b\o_{\m\n}=0. \label{nonDisHom}
\end{align}
Especially, with nonvanishing ${\Th'}_{\d}$ which may significantly modify the dynamical equations of hydrodynamics and bring in extra ambiguity besides pseudogauge and entropy-gauge. For simplicity, we verify in Appendix \eqref{AppConstrt} that the non-dissipative solution of $O\(\pd^0\)$ is unique, which is essentially the leading-order solution in \eqref{decpEm}-\eqref{decpEntr}. It could be interesting to figure out if there are extra non-dissipative solution to $O\(\pd\)$. We leave it for future work.  %, named as non-dissipative-gauge due to its zero entropy production We give the constraints on the non-dissipative-gauge in Appendix \eqref{AppConstrt}. It is tricky to strictly limit the non-dissipative-gauge as zero. We will leave this as future work.  to confirm whether there is non-trivial solution to $\Th_{\d}$ and $s_{\d}$ other than the above two types of transformations. Qualitatively, such kind of solution, if exist, should also be in a `homogeneous' form with factors of arbitrary constants, just as those in entropy-gauge transformation, since the non-semipositive terms as `inhomogeneous' parts have been cancelled out by the solution from pseudogauge transformation.

One can easily verify that the constitutive relations in \eqref{ConRel} give zero entropy production rate in the GTE limit \eqref{GTE}.
\iffalse\begin{align}
0=&\pd_\m s_{\mathrm{GTE}}^{\m}=\[s-\b\(\ep +p - \fr{1}{2}\o_{\m\n} \(\cR^{\m\n} + \cB^{\m\n}\)\)\]\th \label{EntProGTEB} \\
&+\pd_{\m}\(\ms_{\mathrm{GTE}}^{\m} + \mTh_{\mathrm{GTE}}^{\m\n}\b_\n + \fr{1}{2}\mS_{\mathrm{GTE}}\,\e^{\m\a\n\s}\o_{\a\n} u_\s\) .\nonumber
\end{align}\fi
Moreover, keeping only the lowest-order terms in \eqref{EomTR} and taking the separate conservation limits
\begin{align}
\cTh^{[\m\n]}=0,\no
\pd_\a\S^{\a\m\n}+2\Th_{\d}^{[\m\n]}=0, \label{SepConsrv}
\end{align}
in \eqref{EntPro}, we readily confirm that $\pd_\m \(s u^\m\) = 0$. It turns out that the orbital angular momentum conservation in the first equation of \eqref{SepConsrv} contains only the dissipative component of the stress-energy tensor while the spin angular momentum conservation in the second equation have to include the divergence term $\Th_{\d}^{[\m\n]}$. 

\section{Linear-mode analysis}\label{SecLnr}

We perform the linear-mode analysis of the spin hydrodynamic equations \eqref{Em}-\eqref{Am} using the constitutive relations \eqref{ConRel} and \eqref{SolBstGen}. For simplicity, we consider the isotropy background with $O\(\vrp\)\sim O\(\pd\)$. The fluctuations, counted as $O\(\D\)$, are near GTE without background spin density,
\begin{align}
&\ep(x)={\eb}+\ue(x),\qquad p(x)={\pb}+\up(x),\qquad T(x)={\bT}+\uT(x),\no 
&v^i(x)=0+\uv^i(x),\qquad \cR^i(x)=0+\uR^i(x),\qquad \cB^i(x)=0+\uB^i(x), 
\end{align}
with overbar denoting background and underbar denoting fluctuations, where $v^i$ is the fluid three-velocity with $u^\m=(1,v^i)+O(v^2)$. 
Noting $\S_{\d}=O\(\D^2\)$ and using
\begin{align}
\eqref{Euler}\eqref{GibDuh}\quad\to&\quad%\pd_\n\b = \fr{1 }{\ep + p }\[-\b\pd_\n p + \fr{1}{2}\cR_{\a\l} \pd_\n\(\b\o^{\a\l}\)\] ,\no \\ + n \,\pd_\n\(\b\m\) 
\uT = \fr{\bT \up}{\eb+\pb} + O\(\D^2\),
%\eqref{Em}^\m\to\; -\D^{\m\n}\pd_\n p=&\(\ep + p\)D u^\m + u_{\n}D \(h^{\n}+q^{\n}\) u^\m + u_{\a} \pd_\n\(\t^{\n\a}+\ph^{\n\a}\) u^\m\no
% + \th&\(h^\m + q^\m\) + D \(h^\m + q^\m\) + \(h^\n - q^\n\) \pd_\n u^\m + \pd_{\n}\(\t^{\n\m}+\ph^{\n\m}\)
\end{align}
\iffalse
one gets  the identity
\begin{align}
&h^\m-q^\m=-T\k \D^{\m\n}\(D \b_\n-\pd_\n\b\)=O\(\pd^2\).
\end{align}
\fi
we expand \eqref{Em} to $O(\pd^2\D)$ and \eqref{Am} to $O(\pd\D)$ as 
\begin{align}
&\(\pd_0 - c_s^2\k'_s\pd_i\pd^i\)\upi^{0}+\(1+\k'_s \pd_0\)\pd_i\upi^{i} - \(\pd_0 + \G_b\)\pd_i\uB^i=0,\label{emc0}\\ %\eqref{Em}^0\quad\to\quad 0=\pd_0\Th^{00}+\pd_k\(\Th^{0k}+2\mTh_{(a)}^{k0}\)=
%&=\pd_0\d\p^{0}+\pd_k\d\p^{k}+2\pd_k\(\D q^{k}+\d^2 q^{k}+\d\ph^{kj}\d u_j\)\no
& %\eqref{Em}^i\to 0=\pd_0\Th^{0i}+\pd_k\Th^{ki}=\pd_0\upi^{i}+\pd_k\(\D^{ki}\d p+\d\t^{ki}+\d\ph^{ki}\)\no
\pd_0\upi^{i} + c_s^2\pd^{i} \(\upi^0-\pd_k\uB^k\) - \g_{\pr}\pd^{i}\pd_k\upi^{k} - \(\g_{\bot}+\g_s\)\(\d_{k}^{i}\pd_j\pd^{j} - \pd^{i}\pd_k\)\upi^{k}\no
&\myquad[21] - \fr{1}{2}\e^{ijk}\G_r\pd_j\uR_{k}=0,\label{emci}\\
%&\pd_0\upi^{i} + c_s^2\pd^i \upi^0 - \g_{\pr}\pd^i\pd_k\upi^{k} - \(\g_{\bot}+\g_s\)\(\D^{ik}\pd_j\pd^{j} - \pd^{i}\pd_k\)\upi^{k} - \fr{\G_s}{2}\e^{ijk}\pd_j\ubar{\cR}_{k}=0,\no
& \(\pd_0 + \G_b\)\uB^i + \e^{ijk}\pd_k\uR_j + c_s^2\k'_s\pd^i\upi^{0} - \k'_s \pd_0\upi^{i} =0,\label{amc0i}\\ %\e_{ij\m\n}\eqref{Am}^{\m\n}\quad\to\quad
& \pd_0 \uR^i + \G_r \uR^i - 2\g_s\e^{ijk}\pd_j \upi_{k}=0 ,\label{amcij} %\e_{0i\m\n}\eqref{Am}^{\m\n}\quad\to\quad - 2\g_s \pd^j\(\pd_i\uR_j - \pd_j\uR_i \)
\end{align}
\iffalse
\begin{align}
&\(\pd_0 - 2 c_s^2\k'_s\pd_i\pd^i\)\upi^{0}+\(1+2\k'_s \pd_0\)\pd_i\upi^{i} - \(2\G_b+\pd_0\)\pd_i\uB^i - 2 \G_{bR}\pd_i\uR^i=0,\label{emc0}\\ %\eqref{Em}^0\quad\to\quad 0=\pd_0\Th^{00}+\pd_k\(\Th^{0k}+2\mTh_{(a)}^{k0}\)=
%&=\pd_0\d\p^{0}+\pd_k\d\p^{k}+2\pd_k\(\D q^{k}+\d^2 q^{k}+\d\ph^{kj}\d u_j\)\no
& %\eqref{Em}^i\to 0=\pd_0\Th^{0i}+\pd_k\Th^{ki}=\pd_0\upi^{i}+\pd_k\(\D^{ki}\d p+\d\t^{ki}+\d\ph^{ki}\)\no
\pd_0\upi^{i} + c_s^2\pd^{i} \(\upi_0-\pd_k\uB^k\) - \g_{\pr}\pd^{i}\pd_k\upi^{k} - \(\g_{\bot}+\g_s\)\(\d_{k}^{i}\pd_j\pd^{j} - \pd^{i}\pd_k\)\upi^{k}\no
&\myquad[18] - \fr{1}{2}\e^{ijk}\(\G_r\pd_j\uR_{k}+\G_{rB}\pd_j\uB_{k}\)=0,\label{emci}\\
%&\pd_0\upi^{i} + c_s^2\pd^i \upi^0 - \g_{\pr}\pd^i\pd_k\upi^{k} - \(\g_{\bot}+\g_s\)\(\D^{ik}\pd_j\pd^{j} - \pd^{i}\pd_k\)\upi^{k} - \fr{\G_s}{2}\e^{ijk}\pd_j\ubar{\cR}_{k}=0,\no
& \pd_0\uB^i + \e^{ijk}\pd_k\uR_j + 2 c_s^2\k'_s\pd^i\upi^{0} - 2\k'_s \pd_0\upi^{i} + 2\G_b\uB^i + 2 \G_{bR}\uR^i=0,\label{amc0i}\\ %\e_{ij\m\n}\eqref{Am}^{\m\n}\quad\to\quad
& \pd_0 \uR^i + \G_r \uR^i + \G_{rB} \uB^i - 2\g_s\e^{ijk}\pd_j \upi_{k}=0 ,\label{amcij} %\e_{0i\m\n}\eqref{Am}^{\m\n}\quad\to\quad - 2\g_s \pd^j\(\pd_i\uR_j - \pd_j\uR_i \)
\end{align}
\fi
where we have introduced the hydrodynamic and spin modes as
\begin{align}
&\upi^0 \equiv \ubar\Th^{00}=\ue+\pd_i\uB^i+O\(\pd^2\)+O\(\D^2\),\no
&\upi^i \equiv \ubar\Th^{0i}=\(\eb+\pb\)\uv^i-\fr{1}{2}\(\k+\k_s\)\pd_0 \uv^i-\fr{1}{2}\(\k'-\k'_s\)c_s^2\pd^i \ue+\fr{1}{2}\G_b \uB^i+O\(\pd^2\)+O(\D^2) ,\no % +\G_{bR}\uR^i
&\ubar\S^{0ij} = \ubar\cS^{ij} + O\(\pd^2\)+O(\D^2)  = \e^{ijk}\uR_k + O\(\pd^2\)+O(\D^2) ,%\no
%&\ubar\cS^{0i} = \uB^i + O(\D^2) , %+ O\(\pd^2\)
\end{align}
and the constants as
\begin{align}
&c_s^2=\fr{\pd p}{\pd \ep},\quad \g_{\pr}=\fr{1}{\eb+\pb}\(\z+\fr{4}{3}\et\),\quad \g_{\bot}=\fr{\et}{\eb+\pb},\quad 2\g_s=\fr{\et_s}{\eb+\pb},\quad \k'=\fr{\k}{\eb+\pb},\no %,\quad \c_b=-\fr{\pd S^{i0}}{\pd \o^{i0}}
&\c_r\d_j^i=\fr{\pd \cR^{i}}{\pd r^{j}},\qquad \G_r=\fr{2\et_s}{\c_r},\qquad %\c_{rB}\d_j^i=\fr{\pd \cB^{i}}{\pd r^{j}},\qquad \G_{rB}=\fr{2\et_s}{\c_{rB}},\no
\c_b\d_j^i=\fr{\pd \cB^{i}}{\pd b^{j}},\qquad \G_b=\fr{2\k_s}{\c_b} ,\qquad \k'_s=\fr{\k_s}{\eb+\pb}. %\qquad \c_{bR}\d_j^i=\fr{\pd \cR^{i}}{\pd b^{j}},\qquad \G_{bR}=\fr{2\k_s}{\c_{bR}}.
\end{align}
Note that $\upi^0$, $\upi^i$ and $\ubar\S^{0ij}$ are invariant components of $\Th^{\m\n}$ and $\S^{\m\n\a}$ under frame choice \cite{Bhattacharya:2011tra,Kovtun:2012rj}, where $\ubar\S^{0ij}$ can be replaced by $\uR_k$ within linear approximation. The boost modes $\uB^i$ are embedded in the divergence terms of $\Th^{\m\n}$ and can not be defined as the invariant components of spin current since $\ubar\S^{00i}$ vanish for totally antisymmetric $\S^{\m\n\a}$. We have counted $\G_r\sim\G_b\sim 1/\c_r\sim 1/\c_b\sim O\(\pd\)$ in the above linear expansion and neglected the anisotropy in $\fr{\pd \ep}{\pd r}$, $\fr{\pd \ep}{\pd b}$, $\fr{\pd \cR}{\pd b}$ and $\fr{\pd \cB}{\pd r}$. %and $\G_{rB}\sim \G_{bR}\sim 1/\c_{rB}\sim 1/\c_{bR} \sim O\(\pd\D\)$ 
For simplicity, we have taken the speed of sound $c_s$, the susceptibilities $\c_r,\c_b$ and all the kinetic coefficients as constants.

In the Fourier space with $\tcO\(k\)\equiv\int d^4x e^{i\o t-i\bik\cdot \bix}\ubar\cO\(x\)$ and $\bik=\(0,0,k\)$, one finds the block diagonal form of the linearized hydrodynamic equations,
\begin{align}
\(\begin{array}{cccc}
   A_{\pr}^{4\times 4} & O & O & O\\
   O & A_{\bot,\cB}^{2\times 2} & A_+^{2\times 2} & A_-^{2\times 2}\\
%   \hline
   O & O & A_{\bot,+}^{2\times 2} & O\\
%   \hline
   O & O & O & A_{\bot,-}^{2\times 2} 
   \end{array}
\)\vy=0,
\end{align}
with $\vy=\(\upi_0,\upi_z,\uB_z,\uR_z,\uB_x,\uB_y,\upi_x,\uR_y,\upi_y,\uR_x\)^{T}$, where the blocks are % longitudinal and transverse of the matrix 
\begin{align}
A_{\pr}^{4\times 4}&=\(\begin{array}{cccc}
   -i\o+c_s^2\k'_s \bik^2 & i|\bik|+\k'_s\o|\bik| & i\(i\o-\G_b\)|\bik| & 0 \\
   ic_s^2|\bik| & -i\o+\g_\pr\bik^2 & c_s^2 \bik^2 & 0\\
   ic_s^2\k'_s|\bik| & i\k'_s\o & -i\o+\G_b & 0 \\
   0 & 0 & 0 & -i\o+\G_r
   \end{array}\),\no
A_{\bot,\cB}^{2\times 2}&=\(\begin{array}{ccc}
   -i\o+\G_b & 0 \\
   0 & -i\o+\G_b 
   \end{array}\),\quad
A_{\bot,\pm}^{2\times 2}=\(\begin{array}{cc}
   -i\o+\(\g_\bot+\g_s\)\bik^2 & \pm\fr{i}{2}\G_r|\bik|\\
   \mp 2i\g_s|\bik| & -i\o+\G_r %-2\g_s\bik^2 
   \end{array}\) ,\no
A_{+}^{2\times 2}&=\(\begin{array}{cc}
   i\k'_s\o & i|\bik|\\
   0 & 0 
   \end{array}\) ,\qquad
A_{-}^{2\times 2}=\(\begin{array}{cc}
   0 & 0 \\
   i\k'_s\o & - i|\bik|
   \end{array}\).
\end{align}
Note that $\o$ denotes frequency in this section, not to be confused with spin potential $\o_{\m\n}$. The power counting in Fourier space is $\G_r\sim\G_b\sim\o\sim O(\bik)$ where \eqref{emc0}-\eqref{emci} are exact to $O\(\bik^2\)$ while \eqref{amc0i}-\eqref{amcij} are accurate to $O\(\bik\)$. Solving the characteristic equations, det$A_{\pr}^{4\times 4}=0$ and det$A_{\bot}^{2\times 2}=0$, we obtain the dispersion relations,
\begin{align}
&\begin{cases}
\blacklozenge \text{ One pair of sound modes: }\o_{\mathrm{sound}}\(\bik\)=\pm c_s|\bik|-\fr{i}{2}\g_\pr\bik^2 \mp c_s^3\k'_s \fr{k^3}{\G_b}+O\(\bik^3\),\\
\blacklozenge \text{ One longitudinal spin-boost mode: }\o_{\mathrm{spin,b},\pr}=-i\G_b-ic_s^2\k'_s\bik^2 + O\(\bik^2\),\\
\blacklozenge \text{ One longitudinal spin-rotation mode: }\o_{\mathrm{spin,r},\pr}=-i\G_r + O\(\bik^2\),
\end{cases}\label{longt}\\
&\quad\blacklozenge \text{ Two transverse spin-boost modes: }\o_{\mathrm{spin,b},\bot}=-i\G_b+O\(\bik^2\),\label{transB}\\
&\begin{cases}
\blacklozenge \text{ Two shear modes: }\o_{\mathrm{shear}}\(\bik\)=-i\g_\bot\bik^2+O\(\bik^3\),\\
\blacklozenge \text{ Two transverse spin-rotation modes: }\o_{\mathrm{spin,r},\bot}=-i\G_r-i\g_s\bik^2+O\(\bik^2\). %-i\g_s\bik^2
\end{cases}\label{transR}
\end{align}
The dispersion relations of both the hydrodynamic modes and spin modes happen to be the same as the phenomenological formulation\cite{Hattori:2019lfp} to $O(\bik)$. However, to $O(\bik^2)$ the dispersion relations of sound modes and longitudinal spin-boost mode are different. As a comparison to \eqref{longt}, we give the results of the phenomenological formulation as follows
\begin{align}
&\begin{cases}
\blacklozenge \text{ One pair of sound modes: }\o_{\mathrm{sound}}\(\bik\)=\pm c_s|\bik|-\fr{i}{2}\g_\pr\bik^2  \mp 2c_s^3\k'_s \fr{k^3}{\G_b}+O\(\bik^3\),\\
\blacklozenge \text{ One longitudinal spin-boost mode: }\o_{\mathrm{spin,b},\pr}=-i\G_b-3ic_s^2\k'_s\bik^2 + O\(\bik^2\).
\end{cases}\nonumber%\label{transR}
\end{align}
This implies that if one introduces the hydrodynamic and spin modes based on the frame-invariant components of $\Th^{\m\n}$ and $\S^{\m\n\a}$, the dispersion relations will typically differ depending on the specific formulation of spin hydrodynamics.

\section{Summary and Outlook}\label{SecSum}

We have shown that in the canonical formulation of spin hydrodynamics for Dirac fermions featuring a completely antisymmetric spin tensor and a generic spin potential, the stress-energy tensor must be influenced by spin variables at the first order of gradient. Additionally, the inclusion of boost variables is necessary to uphold the entropy principle. 

When boost variables are included, we conduct a linear-mode analysis utilizing the spin hydrodynamic equations derived from the canonical formulation. Upon comparison with the phenomenological formulation, we observe that the dispersion relations of the sound modes and the longitudinal spin-boost mode differ at the second order of gradient.

The violation of the entropy principle in the absence of boost variables is demonstrated with a general antisymmetric spin potential. It is yet to be determined if spin hydrodynamics can be developed solely using the spatial component $r_{\m\n}$ of the spin potential $\o_{\m\n}$ for general rotational fluids with finite thermal vorticity. Furthermore, in the presence of boost variables, instead of opting the constitutive relations of canonical formulation to be related to the pseudogauge transformation of the phenomenological formulation, it would be intriguing to explore if there exist alternative non-dissipative constitutive relations constrained by \eqref{nonDisHom}, and how such constitutive relations would impact the behavior of the hydrodynamic and spin modes. These aspects are left for future investigation.

\begin{acknowledgments}
We thank Xu-Guang Huang for stimulating discussions in several stages of this work. L.X.Y thanks Shi Pu for useful discussions at a workshop, "The 15th QCD Phase Transition and Relativistic Heavy Ion Physics", on Dec.15-19, 2023. L.X.Y is supported by the China Postdoctoral Science Foundation 2023M730707.  
\end{acknowledgments}

\appendix

\section{Completeness of First-Order spin hydrodynamics}\label{AppConstrt}

For completeness, we confirm that there is no $O\(\pd^0\d\)$ non-dissipative solution to \eqref{nonDis}. To this end, we consider the leading-order terms $\Th_{0}$ and $s_{0}$ in \eqref{ConDcp}. %since it is not relevant in the elimination of the first order terms.
We define%explicitly separate the $O\(\pd \o^0\d\)$ and $O\(\pd^0 \o\d\)$ parts in $\Th_{\d}^{\m\n}$ and $ s_{\d}^\m$ as 
\iffalse
\begin{align}
\Th_{\d}^{\a\s} = \Th_{0}^{\a\s} + \Th_{\pd}^{\a\s} + \Th_{\o}^{\a\s\m\n}  \o_{\m\n} ,\myquad[1]
s_{\d}^\a = s_{0}^\a + s_\pd^\a + s_{\o}^{\a\m\n}\b \o_{\m\n} ,  \myquad[1] \S_\d=\S_0
\end{align}
\fi
\begin{align}
&\Th_{0}^{\a\s} \equiv T\sum_{\d=\cR,\cB}\Th_{\d 0}^{\a\s\n}\d_\n,\myquad[1] 
%\Th_{\pd}^{\a\s} \equiv \sum_{\d=\cR,\cB}\Th_{\d 2}^{\a\s\m\n} \d_\n\pd_\m\ln\b + \Th_{\d 3}^{\a\s\m\n\r} \d_\n\pd_\m u_\r + \Th_{\d 4}^{\a\s\m\n}\pd_\m\d_\n,\no
%&\Th_{\o}^{\a\s\m\n} \equiv \sum_{\d=\cR,\cB}\Th_{\d 5}^{\a\s\m\n\r}\d_\r , 
%\myquad[2] \S_0 =\S_{\d 0}^\m\d_\m ,\myquad[2]
s_{0}^\a\equiv \sum_{\d=\cR,\cB}s_{\d 0}^{\a\n}\d_\n , 
%& s_\pd^\a\equiv \sum_{\d=\cR,\cB}\b\(s_{\d 2}^{\a\m\n}\d_\n\pd_\m\ln\b + s_{\d 3}^{\a\s\m\n} \d_\n\pd_\m u_\s + s_{\d 4}^{\a\m\n}\pd_\m\d_\n\),\myquad[1]
%s_{\o}^{\a\m\n} \equiv \sum_{\d=\cR,\cB}s_{\d 5}^{\a\m\n\s}\d_\s ,\nonumber
\end{align}
where $\Th_{\d 0}$ and $s_{\d 0}$ are $O\(\pd^0 \o^0\d^0\)$ coefficients of $\d_\n$ as expressions in terms of $\b$ and $u^\m$. Here we have excluded the dependence on $\d$ in $\Th_{\d 0}$ and $s_{\d 0}$ since the linear dependence on $\d^\n$ has been factored out\footnote{We expect that the leading-order constitutive relations should exhibit linearity that is homogeneous in both the magnitude and direction of $\d^{\m}$, akin to the behavior in thermodynamic relations where extensive quantities and their corresponding density quantities are linearly homogeneous. Therefore, the only viable covariant linear factor of $\d^\m$ is in the form of a four-vector, while the coefficients of $\d^\m$ do not depend on its magnitude or direction.}. %In principle The case for the $ \o_{\m\n}$ dependence of $O\( \o^0\)$ is similar. Here it is sufficient to focus on the $\d^\m$ dependence.%\footnote{} %, by dimension analysis, we have also explicitly factor out $\b$ in order to define Thus, the only (spin) hydrodynamic variables involved in $\Th_{\d a}$ and $s_{\d a}$ is $u_\m$. 
%Moreover, since it is not possible to construct an $O\(\d\)$ scalar to the leading-order using the provided spin hydrodynamic variables, $\S_{\d}$ is supposed to be zero. The spin polarization as an extansive quantity should be linear homogeneous in the thermodynamic relations and the leading-order constitutive relations. 
The single $O\(\pd^0 \o\d\)$ term $\Th_{0}^{\a\s} \o_{\a\s}$ in \eqref{nonDis} should be vanishing for any value of $ \o_{\a\s}$. This gives the constraint 
\begin{align}
\Th_{\d 0}^{[\a\s]\n}\d_\n=0 .\label{cnstrTh0}
\end{align}
The $O\(\pd \o^0\d\)$ terms in \eqref{EntProIr} can be written as
\begin{align}\label{EntPd1}
\pd_\a s_{0}^\a &+ \pd_\a\Th_{0}^{\a\s}\b_\s = \sum_{\d=\cR,\cB}\pd_\a\(s_{\d 0}^{\a\n}\d_\n\) + \pd_\a\(T\Th_{\d 0}^{\a\s\n}\d_\n\) \b_\s\no
\iffalse
=&\sum_{\d=\cR,\cB}\(\pd_\a s_{\d 0}^{\a\n}\d_\n + s_{\d 0}^{\a\n} \pd_\a\d_\n\) + \[\pd_\a\(T\Th_{\d 0}^{\a\s\n}\)\d_\n + T\Th_{\d 0}^{\a\s\n} \pd_\a\d_\n\] \b_\s\no
=&\sum_{\d=\cR,\cB}\(s_{\d 0,\b}^{\a\n}\d_\n\b\pd_\a\ln\b + s_{\d 0,u}^{\a\n,\m}\d_\n\pd_\a u_\m + s_{\d 0, r}^{\a\n,\m\s}\d_\n\pd_\a \o_{\m\s} + s_{\d 0,\hd}^{\a\n,\m}\hd_{\m,\d}^{\,,\s}\d_\n\pd_\a\d_\s + s_{\d 0}^{\a\n} \pd_\a\d_\n\) \no
&+ \[\(T\Th_{\d 0}^{\a\s\n}\)_{,\b}\d_\n\b\pd_\a\ln\b + T\Th_{\d 0,u}^{\a\s\n,\m}\d_\n\pd_\a u_\m + T\Th_{\d 0, r}^{\a\s\n,\m\r}\d_\n\pd_\a \o_{\m\r} + T\Th_{\d 0,\hd}^{\a\s\n,\m}\hd_{\m,\d}^{\,,\r}\d_\n\pd_\a\d_\r + T\Th_{\d 0}^{\a\s\n} \pd_\a\d_\n\] \b_\s\no
=&\sum_{\d=\cR,\cB}\(s_{\d 0,\b}^{\a\n} + \(T\Th_{\d 0}^{\a\s\n}\)_{,\b}\b_\s\)\d_\n\b\pd_\a\ln\b + \(s_{\d 0,u}^{\a\n,\m} + T\Th_{\d 0,u}^{\a\s\n,\m}\b_\s\)\d_\n\pd_\a u_\m\no
&+ \(s_{\d 0, r}^{\a\n,\m\s} + T\Th_{\d 0, r}^{\a\r\n,\m\s}\b_\r\)\d_\n\pd_\a \o_{\m\s} + \(s_{\d 0}^{\a\n} + T\Th_{\d 0}^{\a\s\n} \b_\s + s_{\d 0,\hd}^{\a\s,\m}\hd_{\m,\d}^{\,,\n}\d_\s + T\Th_{\d 0,\hd}^{\a\r\s,\m}\hd_{\m,\d}^{\,,\n}\d_\s\b_\r\) \pd_\a\d_\n\no
\fi
=&\sum_{\d=\cR,\cB}N_{\d 1}^{\a\n}\d_\n\b\pd_\a\ln\b + N_{\d 2}^{\a\n\m}\d_\n\pd_\a u_\m + N_{\d 3}^{\a\n} \pd_\a\d_\n + N_{\d 4}^{\a\n\m\s}\d_\n\pd_\a \o_{\m\s} ,
\end{align}
where
\begin{align}
&N_{\d 1}^{\a\n} \equiv s_{\d 0,\b}^{\a\n} + \(T\Th_{\d 0}^{\a\s\n}\)_{,\b}\b_\s ,&& N_{\d 2}^{\a\n\m} \equiv s_{\d 0,u}^{\a\n,\m} + \Th_{\d 0,u}^{\a\s\n,\m}u_\s ,\\
%&+ s_{\d 0,\hd}^{\a\s,\m}\hd_{\m,\d}^{\,,\n}\d_\s + T\Th_{\d 0,\hd}^{\a\r\s,\m}\hd_{\m,\d}^{\,,\n}\d_\s\b_\r&& aa\no
&N_{\d 3}^{\a\n} \equiv s_{\d 0}^{\a\n} + \Th_{\d 0}^{\a\s\n} u_\s ,&& N_{\d 4}^{\a\n\m\s} \equiv s_{\d 0, \o}^{\a\n,\m\s} + T\Th_{\d 0, \o}^{\a\r\n,\m\s}\b_\r ,\nonumber % 
\end{align}
with notations $A_{,\b}\equiv\pd A/\pd \b$, $A_{,u}^{,\m}\equiv\pd A/\pd u_\m$ and $A_{, \o}^{,\m\n}\equiv\pd A/\pd  \o_{\m\n}$. %$A_{,\hd}^{,\m}\equiv\pd A/\pd \hd_\m$ and $\hd_{\m,\d}^{\,,\n}\equiv\pd\hd_\m/\d_\n$. 
Noting $u\cdot u=-1$ and $u\cdot\d=0$, \eqref{EntPd1} should be identically zero in groups as follows,
\begin{align}
&N_{\d 1}^{\a\n}\d_\n\b\pd_\a\ln\b=0,\no
&N_{\d 2}^{\a\n\m}\d_\n\pd_\a u_\m + N_{\d 3}^{\a\n} \pd_\a\d_\n = 0 = M_{\d 0}^{\a\n}\d_\n\pd_\a\(u\cdot u\) + M_{\d 2}^{\a}\pd_\a\(u\cdot\d\) , 
\end{align}
where the constraints can be written as
\begin{align}
&N_{\d 1}^{\a\n}\d_\n = 0 \;\text{ and }\; \cN_{\d 2}^{\a\n\m}\d_\n = 0 \;\text{ and }\; \cN_{\d 3}^{\a\n} = 0 \;\text{ and }\; \cN_{\d 4}^{\a\n[\m\s]}\d_\n = 0 ,\label{cnstrThS0}
\end{align}
with 
\begin{align}
&\cN_{\d 2}^{\a\n\m} \equiv N_{\d 2}^{\a\n\m} - 2M_{\d 0}^{\a\n} u^\m - M_{\d 2}^{\a}g^{\n\m} ,  
\myquad[3] \cN_{\d 3}^{\a\n} \equiv N_{\d 3}^{\a\n}-M_{\d 2}^{\a} u^\n .
\end{align}
Using the above constraints we have
\begin{align}
&\cN_{\d 3}^{\a\n} = 0 \;\to\; 0=\cN_{\d 3}^{\a\n}\d_\n = N_{\d 3}^{\a\n}\d_\n\;\to\; 0=\(N_{\d 3}^{\a\n}\d_\n\)_{,u}^{,\m}=\(N_{\d 2}^{\a\n\m}+\Th_{\d 0}^{\a\m\n}\)\d_\n ,\no
&\cN_{\d 2}^{\a\n\m}\d_\n = 0 \;\to\; 0=\(\Th_{\d 0}^{\a\m\n}+2M_{\d 0}^{\a\n} u^\m + M_{\d 2}^{\a}g^{\n\m}\)\d_\n\;\to\; 0=\(\Th_{\d 0}^{\a\m\n}u_\m - 2M_{\d 0}^{\a\n}\)\d_\n,\no
&0=\(N_{\d 3}^{\a\n}\d_\n\)_{,\b}=\(N_{\d 1}^{\a\n}+T\Th_{\d 0}^{\a\s\n}u_\s\)\d_\n=T\Th_{\d 0}^{\a\s\n}u_\s\d_\n\;\to\; \Th_{\d 0}^{\a\s\n}u_\s\d_\n=0 ,\no
&0=N_{\d 3}^{\a\n}\d_\n=\(s_{\d 0}^{\a\n} + \Th_{\d 0}^{\a\s\n} u_\s\)\d_\n\;\to\; s_{\d 0}^{\a\n}\d_\n=0 ,\\
&\cN_{\d 2}^{\a\n\m}\d_\n u_\m = 0 \;\to\; 0=\(\Th_{\d 0}^{\a\m\n}u_\m - 2M_{\d 0}^{\a\n}\)\d_\n=- 2M_{\d 0}^{\a\n}\d_\n\;\to\; M_{\d 0}^{\a\n}\d_\n=0 ,\no
&0=\(\Th_{\d 0}^{\a\m\n}+2M_{\d 0}^{\a\n} u^\m + M_{\d 2}^{\a}g^{\n\m}\)\d_\n=\Th_{\d 0}^{\a\m\n}\d_\n + M_{\d 2}^{\a}\d^\m\;\to\; \Th_{\d 0}^{\a\m\n}\d_\n =- M_{\d 2}^{\a}\d^\m . \no
&\cN_{\d 2}^{\a\n\m}\d_\n = 0 \;\to\; 0=N_{\d 2}^{\a\n\m}\d_\n - M_{\d 2}^{\a}\d^{\m} = \(s_{\d 0}^{\a\n}\d_\n\)_{,u}^{,\m} + \(\Th_{\d 0}^{\a\s\n}\d_\n\)_{,u}^{,\m}u_\s - M_{\d 2}^{\a}\d^{\m}= - M_{\d 2}^{\a}\d^{\m},\nonumber
\end{align}
%Given $\Th_{\d 0}^{[\a\s]\n}\d_\n=0$, we get the constraint $M_{\d 2}^{\a}=\cM_{\d 2}\hd^\a$ with $\hd^\a\equiv \d^\a/|\d|$ where $\cM_{\d 2}$ could be any possible dimensionless scalar as expressions in terms of $\b,u, r$ and $\d$. Subbing $\Th_{\d 0}^{\a\s\n}\d_\n=\cM_{\d 2}\hd^\a\d^\s$ in $\cN_{\d 2}^{\a\n\m}\d_\n = 0$, we obtain
which gives $M_{\d 2}^\a=0$. Consequently, the combined constraints from \eqref{cnstrTh0} and \eqref{cnstrThS0} lead to $s_{0}^\a =0$ and $\Th_{0}^{\a\s}=0$. This means that there are no other zeroth-order non-dissipative terms in \eqref{decpEm}-\eqref{decpEntr}.

\bibliographystyle{unsrt}
\bibliography{hydro_eqns.bib}
 
\end{document}